\documentclass{article}
\usepackage[a4paper, portrait, margin=1.1811in]{geometry}
\usepackage[english]{babel}
\usepackage[utf8]{inputenc}
\usepackage[T1]{fontenc}
\usepackage{helvet}
\usepackage{etoolbox}
\usepackage{graphicx}
\usepackage{titlesec}
\usepackage{caption}
\usepackage{booktabs}
\usepackage{xcolor} 
\usepackage[colorlinks, citecolor=cyan]{hyperref}
\usepackage{caption}
\graphicspath{ {./images/} }
\usepackage{scrextend}
\usepackage{fancyhdr}
\usepackage{graphicx}
\usepackage{amsthm}
\usepackage{tikz}
\usepackage{caption}
\usepackage{subcaption}
\usepackage{soul}
\usepackage{tikz-cd, amssymb, amsmath}
\usepackage{tikz}
\usetikzlibrary{shapes,arrows,chains}
\usetikzlibrary[calc]
\usepackage{verbatim}

 \usepackage{mathtools}
 \usepackage{amsmath,amsfonts,amssymb,amsthm,epsfig,epstopdf,url,array}
\usepackage{graphicx}

\usepackage{amsmath} 
\usepackage{amssymb} 
\usepackage[english]{babel}
\usepackage{subcaption}
\usepackage{float}
\usepackage{here}
\usepackage[all]{xy}
\usepackage{mathrsfs}
\usepackage[scr=rsfs,cal=boondox]{mathalfa}

\newtheorem{theorem}{Theorem}
\newtheorem{lema}{lemma}
\newtheorem{prop}{Proposition}
\newtheorem{defi}{Definition}
\newtheorem{remark}{Remark}
\newtheorem{conjecture}{Conjecture}

\newtheorem{lemma}[lema]{Lemma}
\newtheorem{corollary}{Corollary}

\newtheorem{proposition}[prop]{Proposition}

\newtheorem{definition}[defi]{Definition}

\usepackage{comment}
\usepackage{bbold}
\usepackage{pgfplots}
\pgfplotsset{compat=1.11}
\usepgfplotslibrary{fillbetween}
\usetikzlibrary{intersections}
\pgfdeclarelayer{bg}
\pgfsetlayers{bg,main}

\usepackage{tocloft}

\cftsetindents{section}{0em}{2em}
\cftsetindents{subsection}{0em}{2em}

\usepackage{ulem}

\makeatletter
\patchcmd{\@maketitle}{\LARGE \@title}{\fontsize{16}{19.2}\selectfont\@title}{}{}
\makeatother

\newcommand{\be}{\begin{equation}}
\newcommand{\ee}{\end{equation}}

\newcommand{\Hilbert}{\mathscr{H}}

\DeclareMathOperator{\tr}{tr}

\DeclareMathOperator{\rk}{rank}

\usepackage{authblk}

\newsavebox\affbox
\author[1,2]{\textbf{Pablo Costa Rico}}
\affil[1]{Department of Mathematics, Technische Universität München, Germany.
}
\affil[2]{Munich Center for Quantum Science and Technology (MCQST), Germany.}

\titlespacing\section{0pt}{12pt plus 4pt minus 2pt}{0pt plus 2pt minus 2pt}
\titlespacing\subsection{12pt}{12pt plus 4pt minus 2pt}{0pt plus 2pt minus 2pt}
\titlespacing\subsubsection{12pt}{12pt plus 4pt minus 2pt}{0pt plus 2pt minus 2pt}

\titleformat{\section}{\normalfont\fontsize{10}{15}\bfseries}{\thesection.}{1em}{}
\titleformat{\subsection}{\normalfont\fontsize{10}{15}\bfseries}{\thesubsection.}{1em}{}
\titleformat{\subsubsection}{\normalfont\fontsize{10}{15}\bfseries}{\thesubsubsection.}{1em}{}

\titleformat{\author}{\normalfont\fontsize{10}{15}\bfseries}{\thesection}{1em}{}

\title{\textbf{\huge New Partial Trace Inequalities and Distillability of Werner States}\\
	}
\date{}    

\begin{document}

\pagestyle{headings}	
\newpage
\setcounter{page}{1}
\renewcommand{\thepage}{\arabic{page}}

\newcommand{\PCR}[1]{{\color{blue}#1}}
	
\captionsetup[figure]{labelfont={bf},labelformat={default},labelsep=period,name={Figure }}	\captionsetup[table]{labelfont={bf},labelformat={default},labelsep=period,name={Table }}
\setlength{\parskip}{0.5em}
	
\maketitle
	
\noindent\rule{15cm}{0.5pt}
	\begin{abstract}
        One of the oldest problems in quantum information theory is to study   if there exists a state with negative partial transpose which is undistillable \cite{Problem}. This problem has been open for almost 30 years, and still no one  has been able to give a complete answer to it.  This work presents a new strategy to try to solve this problem by translating  the distillability condition on the family of Werner states  into a problem of partial trace inequalities, this is the aim of our first main result.  As a consequence we obtain a new bound for the $2$-distillability of Werner states, which does not depend on the dimension of the system.  On the other hand, our second main result provides new partial trace inequalities for bipartite systems, connecting some of them also with the separability of Werner states.   Throughout this work we also present numerous  partial trace inequalities, which are valid for many families of matrices.

		\textbf{\textit{Keywords}}: \textit{Werner states; Distillability; Partial trace; Bound entanglement; Trace inequalities.}
	\end{abstract}
\noindent\rule{15cm}{0.4pt}

\let\thefootnote\relax\footnotetext{
			 \hspace{-15pt}\small $^{*}$\textbf{Pablo Costa Rico} \textit{
				\textit{E-mail address: \color{cyan}pablo.costa@tum.de}}\\
		}

{
  \hypersetup{linkcolor=black}
}

\section{Introduction}

The theory of quantum entanglement, introduced in \cite{Einstein} in 1935 by Einstein, Podolsky and Rosen,  has been one of the central topics of debate and progress in the last century in quantum mechanics. However, many questions remain to be solved in this field, and in this work, we will discuss one of them, the famous problem stated in \cite{Problem}: Study if  there exists a state with negative partial transpose which is undistillable.
 A quantum state $\rho \in L(\Hilbert)$, where $\Hilbert=\mathcal{H}_1 \otimes \mathcal{H}_2$,  is a positive semidefinite matrix with $\tr \rho=1$. A state is called separable if it can be written as a convex sum of  tensor products of positive semidefinite matrices. Otherwise, it is called entangled. 

In this paper, we will focus on the fundamental property called distillability: Suppose that we have two parties, call them Alice and Bob, who share $n$-copies of the same state  $\rho \in L(\mathbb{C}^d \otimes \mathbb{C}^{d})$, $\rho \geq 0$, $\tr \rho=1$, and that both perform a local operation obtaining a new state of the form 
    \begin{equation}\label{rhoprima}
        \rho'=\frac{A \otimes B\rho^{\otimes n}A^* \otimes B^*}{\tr[A \otimes B\rho^{\otimes n}A^* \otimes B^*]},
    \end{equation}
with $A,B: (\mathbb{C}^d)^{\otimes n} \to \mathbb{C}^2$, and where "*" denotes the adjoint matrix.  If  it is possible to find a pair of operations $(A,B)$ such that the resulting state $\rho'$ is entangled,  it is said that $\rho$ is n-distillable (see e.g. \cite{Horodecki}) . If, on the other hand, for any pair of operations $(A,B)$ the state $\rho'$ is always separable, we say that  $\rho$ is $n$-undistillable.  If for every $n \in \mathbb{N}$, $\rho$ is $n$-undistillable, then $\rho$ is called simply undistillable, otherwise it is distillable. An alternative definition is that $\rho$ is $n$-undistillable if, for every Schmidt rank $2$ vector $v \in (\mathbb{C}^d\otimes\mathbb{C}^d)^{\otimes n}$,
 \begin{equation}\label{AlternativeDefDist}
     \langle v, \left(\rho^{T_1}\right)^{\otimes n}  v \rangle \geq 0,
 \end{equation}
 where $T_1$ denotes the partial transposition and with the Schmidt rank  defined as the minimum
number of terms needed to express a quantum state as
a sum of tensor  product states, see \cite{Lewenstein}, \cite{Pankowski} or \cite{Dur}.

 In \cite{HorodeckiWerner}, it was shown that it is enough to reduce the distillability problem to the family of  Werner states  defined  as (see e.g. \cite{Pankowski} or \cite{Vianna})
    \begin{equation}\label{def:WernerState}
        \rho_{\alpha}=\frac{\mathbb{1}+\alpha F}{d^2+\alpha d},
    \end{equation}
where $\alpha \in [-1,1]$ and  $F$ is the flip operator acting on tensor products as $F(x \otimes y)=y \otimes x$, for $x,y \in \mathbb{C}^d\otimes \mathbb{C}^d$. The reason for that, is that every state with a non-positive partial transpose  can be mapped onto a Werner state with a non-positive partial transpose via the twirling map
\begin{equation}
        \mathcal{T}(\rho)=\int_{U(d)} (U \otimes U)\rho(U \otimes U)^* dU,
    \end{equation}
where $d U$ denotes the Haar measure on the unitary group of $d\times d$ matrices $U(d)$. This twirling operator is in particular a form of local operations and classical communication (LOCC) since it consists of
a convex combination of local unitary operators. Therefore, the existence of undistillable states with non-positive partial transpose can be decided just by focusing on Werner states. This family of states satisfy the following properties:
\begin{itemize}
    \item[1.] $\rho_{\alpha}$ is separable $\Leftrightarrow$  $\rho_{\alpha} \text{ has positive partial transpose}$ $\Leftrightarrow$  $\alpha \geq -\frac{1}{d}$.
    \item[2.] For $n=1$ in \eqref{rhoprima}, $\rho_{\alpha}$ is $1$-undistillable  $ \Leftrightarrow \alpha \geq -\frac{1}{2}$. 
\end{itemize}
Moreover, in  \cite{Horodecki5Problems} it is conjectured that this family might contain a subfamily of states which are undistillable but with non-positive partial transpose. In the last decades, there have been many different approaches to this problem, some of them leading to particular results that have been proved for the distillability of the Werner states , for example in \cite{Chen1}, \cite{Chen2}, \cite{Lewenstein}, \cite{Pankowski}, but still the problem remains open. The conjecture on this family of Werner states is the following:

\begin{conjecture}\label{conjecture1}
   Let $\alpha \in [-1,1]$, $d\geq 2$.  A Werner state $\rho_{\alpha} \in L(\mathbb{C}^d \otimes \mathbb{C}^d)$,  is $n$-undistillable for every $n \in \mathbb{N}$ if, and only if $\alpha \geq -\frac{1}{2}$. 
\end{conjecture}

A positive answer to  Conjecture \ref{conjecture1} would solve then the problem of finding a state with negative partial transpose and is undistillable.

\subsection{Summary of main results and structure of this work}

In this work we provide a new  characterization for the Conjecture \ref{conjecture1} in terms of   partial trace inequalities for the 2-norm, which depend on the  parameter $\alpha$ associated to the Werner states \eqref{def:WernerState}. This is the goal of the first main result, Theorem \ref{Theorem:2distillability}, which is presented  in section  \ref{sec:distillability}. Moreover in Proposition \ref{Prop:DimensionBound} we also show  the connection between the separability and another family of partial trace inequalities. The following map shows the connection:

\tikzstyle{block} = [rectangle, draw, text width=7em, text centered, rounded corners, minimum height=4em]
\tikzstyle{line} = [draw, <->]

\begin{figure}[H]
\begin{center}
\begin{tikzpicture}[node distance=1cm, auto]
\node (init) {};
\node (E)[draw] at (0,4.2) {Werner states};
\node (F)[draw] at (7,4.2) {Partial trace inequalities };

\node at (0,2) [block] (A) {Distillability};
\node at (0,2) [block, right=4cm of A] (B) {Family of partial trace inequalities 1};

\path [line] (A) -- node [text width=2.5cm,midway,above ] {Theorem \ref{Theorem:2distillability}} (B);
 \draw (0,1) ellipse [x radius=55pt, y radius=80pt];
 \draw (6.7,1) ellipse [x radius=55pt, y radius=80pt];

\node at (0,0)[block] (C) {Separability};
\node (0,0) [block, right=4cm of C] (D) {Family of partial trace inequalities 2};

\path [line] (C) -- node [text width=2.5cm,midway,above ] {Proposition \ref{Prop:DimensionBound}} (D);
\end{tikzpicture}
\end{center}
\end{figure}
In section \ref{sec:PartialTraceInequalities}, we will study how the quadratic forms associated to  state inversion operators studied in e.g. \cite{Huber}, \cite{Eltschka} or \cite{Lewenstein2} are related with the distillability and separability properties of Werner states and tensor product of Werner states.  We will exploit the correspondence shown in the previous figure to obtain new results on both partial trace inequalities and properties of Werner states.  In Proposition \ref{Prop:DimensionBound}, we will use that $\rho_{\alpha}$ is separable  for $\alpha \geq -\frac{1}{d}$ to prove partial trace inequalities in $n$-partite systems for arbitrary matrices.

 Theorem \ref{theo:inequalities} is our second main result, where we present the partial trace inequalities that we have been able to prove for bipartite systems. The spirit  will be then the opposite as before, that is, we will try to  prove  partial trace inequalities to  obtain information on the $2$-distillability properties of Werner states. In Theorem \ref{theo:inequalities} appear four partial trace inequalities: One in the family 1 related to the distillability, the second one related to the family 2, and two others more concerning the distillability of the tensor product of two  Werner states with different sign in the parameter $\alpha$ (see Remark \ref{rem:PlusMinus}).  In particular, we  show that both $\rho_{\frac{1}{2}}\otimes \rho_{-\frac{1}{2}}$  and $\rho_{-\frac{1}{2}}\otimes \rho_{\frac{1}{2}}$ are $1$-distillable. The proof of Theorem \ref{theo:inequalities} is presented in  Section \ref{sec:ProofTheorem2}, due to its length. 
 
 Another remarkable result is Corollary \ref{Coro1/4} in section \ref{sec:distillability}. There, we prove that for $\alpha \geq -\frac{1}{4}$, the Werner states $\rho_{\alpha}$   are $2$-undistillable for every dimension $d\geq 2$. This becomes relevant for $d\geq 5$, since then  for $\alpha \in (-\frac{1}{d},-\frac{1}{4}]$ this implies that  the states $\rho_{\alpha}$  are $2$-undistillable and entangled.
 
  In section \ref{sec:3Hilbert}, we discuss about partial trace inequalities in tripartite systems, and prove a particular case of the quadratic form associated to the $3$-distillabilty. Finally in section \ref{sec:NumericalResults}, we present numerical results showing the existence of general families of partial trace inequalities for all the Schatten $p$-norms. 

\section{Preliminaries}
Let $\Hilbert$ be a finite-dimensional Hilbert space. We will denote the set of bounded linear operators in $\Hilbert$ by $L(\Hilbert)$. For $T \in L(\Hilbert)$ the Schatten $p$-norms are defined for $p>0$ as
\begin{equation}
    \Vert T \Vert_p=\left(\tr\vert T \vert^p \right)^{\frac{1}{p}},
\end{equation}
where $\vert T \vert=\sqrt{T T^*}$. In particular, for $p=2$, this norm comes from an inner product in $L(\Hilbert)$ called Hilbert-Schmidt product defined as
\begin{equation}
    \langle T, S \rangle =\tr(T^*S),
\end{equation}
for $T,S \in L(\Hilbert)$. The case $p=\infty$ corresponds with the operator norm. In the particular case where $\Hilbert=\Hilbert_1\otimes \Hilbert_2$, one can define the partial trace operator valued functions for $T \in L(\Hilbert_1 \otimes \Hilbert_2)$, $T=\displaystyle \sum_{i=1}^n T_i^1\otimes T_i^2$ as
\begin{equation}
    \tr_1 T=\sum_{i=1}^n \tr(T_i^1)T_i^2, \qquad \tr_2 T=\sum_{i=1}^n \tr(T_i^2)T_i^1,
\end{equation}
which are independent from the choice of decomposition in tensor products. The following inequalities show some well-known bounds for the norms $1$ and $2$
\begin{equation}\label{Inequality12norm}
    \Vert T \Vert_2 \leq \Vert T \Vert_1 \leq \sqrt{r} \Vert T \Vert_2,
\end{equation}
\begin{equation}\label{RelationNormTrace}
    \Vert T \Vert_2^2\geq \frac{1}{r}\vert \tr T \vert^2,
\end{equation}
\begin{equation}\label{InequalityPartialTrace1}
    \Vert \tr_i T \Vert_1 \leq \Vert T  \Vert_1
\end{equation} 
for $i=1,2$ and where $r=\rk(T)$. For \eqref{InequalityPartialTrace1} see e.g. \cite{Rastegin}.

\begin{remark}
    Many times in the literature in quantum mechanics, the bra-ket notation is used for normalized vectors. In this work however, we will use this notation to denote arbitrary rank 1 matrices, i.e. matrices of the form $\vert v \rangle \langle w \vert$ where $v,w \in \Hilbert$, but not vectors or associated functionals. There will be only one exception for this in the proof of  Proposition \ref{theo:PartialTraceCraeationAnnihilation}, where it simplifies the notation, and it is also indicated there.
\end{remark}

 Given a Hilbert space $\mathcal{H}$, the symmetric and antisymmetric subspaces for two copies of $\Hilbert$ are defined as
\begin{equation}
    \Hilbert_{+}=\{v \in \Hilbert \otimes \Hilbert: Fv=v\}, \quad \Hilbert_{-}=\{v \in \Hilbert \otimes \Hilbert: Fv=-v\},
\end{equation}
respectively, where $F$ is the flip operator. The respective orthogonal projections are given by
\begin{equation}
    P_+=\frac{\mathbb{1}+F}{2}, \quad P_-=\frac{\mathbb{1}-F}{2}.
\end{equation}
For $v,w \in  \Hilbert$, define the symmetric product $\odot:\Hilbert \otimes\Hilbert \to \Hilbert_+$ and the antisymmetric product $\wedge:\Hilbert \otimes\Hilbert  \to \Hilbert_-$
\begin{equation}
   v \odot w=v \otimes w+w\otimes v, \quad  v \wedge w=v \otimes w-w\otimes v.
\end{equation}
Finally, the bosonic and fermionic creation  operators acting on $w \in \Hilbert$ are 
\begin{equation}\label{def:creation}
    a_+^*(v)w=\sqrt{2}P_+(v\otimes w)=\frac{1}{\sqrt{2}}(v \odot w), \quad a_-^*(v)w=\sqrt{2}P_-(v \otimes w)=\frac{1}{\sqrt{2}}(v \wedge w),
\end{equation}
respectively, for $v \in \Hilbert$, and the bosonic and fermionic annihilation  operators on $\varphi \in \Hilbert\otimes \Hilbert$ are just
\begin{equation}\label{def:annihilation}
    a_+(v)(\varphi)=\sqrt{2}\langle v, P_+\varphi \rangle_1, \quad  a_-(v)(\varphi)=\sqrt{2}\langle v, P_-\varphi \rangle_1,
\end{equation}
where $\langle \hspace{3pt}, \hspace{3pt} \rangle_1: \Hilbert \times \Hilbert^{\otimes 2} \to   \Hilbert$ is the partial inner product in the first argument, i.e., the sesquilinear extension of 
\begin{equation}\label{innerannihilation}
(v, \varphi_1 \otimes \varphi_2)  \mapsto  \langle v, \varphi_1 \rangle \varphi_2.
\end{equation}
See \cite{Bratteli} for a more general definition of the creation and annihilation operators in the Fock space.

\section{Distillability of Werner states}\label{sec:distillability}

In recent times, on the way to solving the Werner states' distillability problem, several equivalent problems have been proposed in order to approach it with different strategies. A first example is the one  formulated in \cite{Pankowski} for the  $\mathbb{C}^4 \otimes \mathbb{C}^4$ system, where the 2-distillability problem is equivalent to show that for matrices $A,B \in L(\mathbb{C}^4)$, with $\tr A=\tr B=0$ and $\Vert A \Vert_2^2+\Vert B \Vert_2^2=\frac{1}{2}$,  the two largest singular values squared of the Kronecker sum $A \otimes \mathbb{1}+\mathbb{1}\otimes B $ are upper bounded by $\frac{1}{2}$. A second one is provided in \cite{DiVincenzo} and it relates  the undistillability with the  existence of  a completely positive map, which is not completely co-positive but 2-copositive in all its tensor $n$-th tensor power, see also \cite{Horodecki5Problems}. Our first main result  presents a new characterization in terms of partial trace inequalities.

\begin{theorem}\label{Theorem:2distillability}
Let $\Hilbert$ be a finite-dimensional Hilbert space that can be decomposed as $\Hilbert=\Hilbert_1\otimes \hdots \otimes \Hilbert_n$, with $\dim (\Hilbert_i)=d_i$, and define for $\alpha \in \mathbb{R}$ the quadratic form
\begin{equation}\label{Definitionqn}
    q^{(n)}(\alpha,C)= \sum_{J \in P(\{1,2, \hdots ,n\})  }\alpha^{\vert J \vert}\Vert \tr_{J} C \Vert_2^2,
\end{equation}
where $P(X)$ is the power set of X and we  denote $tr_{\emptyset} = \mathbb{1}$.  Then, $\rho_{\alpha}$ is $n$-distillable  if and only if there exists a matrix  $C \in L((\mathbb{C}^d)^{\otimes n})$ with rank $C \leq 2$ such that $q^{(n)}(\alpha,C) <0$, with $\alpha \in[-1,1]$.
\end{theorem}
\begin{proof}
Suppose that $\rho_{\alpha}$ is $n$-copies distillable, i.e,  there exists $A,B$ such that $\rho' \in L(\mathbb{C}^2\otimes \mathbb{C}^2)$ in \eqref{rhoprima} is entangled. Since the Hilbert space for $\rho'$ is $\mathbb{C}^2\otimes \mathbb{C}^2$, this implies (see  \cite{HorodeckiC6}) that $(\rho')^{T_1} \ngeq 0$, so there exists an element $\psi \in \mathbb{C}^2 \otimes \mathbb{C}^2$ such that
\begin{equation}\label{PropMichaelMenor0}
\langle \psi, (\rho')^{T_1} \psi \rangle<0.
\end{equation}
Let $V \in L(\mathbb{C}^2)$ such that $\psi=(\mathbb{1} \otimes V^* ) \Omega$, where $\Omega$ is the maximally entangled state and denote by $P_{\Omega}$ the orthogonal projection onto $\Omega$, and by $F^{\mathbb{C}^{2}}$ the flip operator in $\mathbb{C}^2\otimes \mathbb{C}^2$, which satisfy the relation $P_{\Omega}^{T_1}=\frac{1}{2}F^{\mathbb{C}^{2}}.$  We can then write
\begin{subequations}
\begin{align}
    \langle \psi, (\rho')^{T_1} \psi \rangle &=  \tr[P_{\Omega}(\mathbb{1} \otimes V)(\rho')^{T_1}(\mathbb{1} \otimes V)^*] \\
    &=\tr[P_{\Omega}^{T_1}(\mathbb{1} \otimes V)(\rho')(\mathbb{1} \otimes V)^*] \\
    &=\frac{1}{2}\tr[F^{\mathbb{C}^{2}}(\mathbb{1} \otimes V)(\rho')(\mathbb{1} \otimes V)^*] \\
    &\backsim \tr[F^{\mathbb{C}^{2}}(A \otimes VB)(\rho_{A_1B_1} \otimes  \hdots \otimes \rho_{A_nB_n})(A \otimes VB)^*], \label{eq:subeq2}
\end{align}
\end{subequations}
 where $\backsim$ means up to the normalization factor. Defining $D=VB$ and using the cyclical property of the trace,
\begin{equation}
    \tr\left[(\rho_{A_1B_1}\otimes \hdots \otimes \rho_{A_nB_n})(A \otimes D)^*F^{\mathbb{C}^2}(A \otimes D)\right]<0.
\end{equation}
Let $\tilde{F}$ be the linear extension of the operator which acts on the tensor product as
$
\tilde{F}(x \otimes y)=y \otimes x,$ $ x,y \in (\mathbb{C}^d)^{\otimes n}.
$
Then,
$$
(A \otimes D)^*F^{\mathbb{C}^2}(A \otimes D)(x \otimes y)=(A^*Dy)\otimes(D^*Ax)=(A^*D\otimes D^*A)(y \otimes x),
$$
so we obtain the relation
\begin{equation}\label{Flipseq}
(A \otimes D)^*F^{\mathbb{C}^2}(A \otimes D)=(A^*D\otimes D^*A)\tilde{F}.
\end{equation}

Now, let  $C=A^*D$. On the one hand,  this matrix satisfies that
$$
\rk(C)=\rk(A^*BV)\leq \min\{ \rk(V), \rk(A),\rk(B)\}\leq 2.
$$
On the other hand, denote by $P(\{1,2, \hdots, n\})$  the power set of $\{1,2, \hdots, n\}$ and define for $J \in P(\{1,2, \hdots, n\}) $, $F_{A_JB_J}$ to be the flip operator swapping the systems $A_j$ and $B_j$ for every $j\in J$. Then if $J^C$ is the complementary set of $J$ in $P(\{1,2, \hdots, n\})$, combining \eqref{eq:subeq2} and \eqref{Flipseq},
\begin{subequations}
    \begin{align}
        \langle \psi, (\rho')^{T_1} \psi \rangle &\backsim \tr\left[\left(\mathbb{1}+\alpha F_{A_1B_1}\right)\otimes \hdots \otimes\left(\mathbb{1}+\alpha F_{A_nB_n}\right)(C \otimes C^*) \tilde{F} \right]\\
        &=\sum_{J \in P(\{1,2, \hdots, n\})}\alpha^{\vert J \vert} \tr\left[ F_{A_JB_J}(C \otimes C^*)\tilde{F} \right]\\
        &=\sum_{J \in P(\{1,2, \hdots, n\})}\alpha^{\vert J \vert} \tr \left[(C \otimes C^*) F_{A_{J^C}B_{J^C}}\right]\\
        &= \sum_{J \in P(\{1,2, \hdots, n\})}\alpha^{\vert J \vert} \tr_{A_{J^C}B_{J^C}} \left[\tr_{A_JB_J}(C \otimes C^*) F_{A_{J^C}B_{J^C}}\right]\\
         &=\sum_{J \in P(\{1,2, \hdots, n\})}\alpha^{\vert J \vert} \tr_{A_{J^C}B_{J^C}} \left[(\tr_J C \otimes \tr_JC^*) F_{A_{J^C}B_{J^C}}\right]\\
          &=\sum_{J \in P(\{1,2, \hdots, n\})}\alpha^{\vert J \vert} \Vert \tr_J C\Vert_2^2,
    \end{align}
\end{subequations}
where in the last equation we used the "swap  trick"  $\tr_{A_{J^C}B_{J^C}}[(X \otimes Y)F_{A_{J^C}B_{J^C}}]=\tr_{A_{J^C}B_{J^C}}[XY]$.

Conversely, suppose that there exists a matrix $C\in L((\mathbb{C}^d)^{\otimes n})$ with rank lower or equal than 2 such that $q^{(n)}(\alpha,C)<0$. By the previous argument this  implies that 
\begin{equation}\label{PropoMichaelDesigualdad}
\tr[\rho_{A_1B_1} \otimes \hdots \otimes \rho_{A_nB_n}(C \otimes C^*)\tilde{F}]<0.
\end{equation}
Decompose $C=\vert v_1 \rangle \langle w_1 \vert+\vert v_2 \rangle \langle w_2 \vert$, and notice that
   \begin{equation}
       (C \otimes C^* \tilde{F})^{\tilde{T}_1}=\vert \psi_C \rangle \langle \psi_C \vert,
   \end{equation}
   where $ \psi_C = v_1\otimes  w_1 + v_2 \otimes  w_2 $ and following  the notation for the flip operator $\tilde{F}$, we denote $ (\hspace{2pt} \cdot \hspace{2pt})^{\tilde{T}_1}$  the partial transposition for  $(\mathbb{C}^d)^{\otimes n}\otimes (\mathbb{C}^d)^{\otimes n} $. Then,
   \begin{equation}
       q^{(n)}(\alpha,C)=\langle \psi_C,(\mathbb{1}+ \alpha F_{A_1B_1})^{T_1}\otimes \hdots \otimes (\mathbb{1}+ \alpha F_{A_nB_n})^{T_1}\psi_C \rangle<0,
   \end{equation}
   and hence, $\rho_{\alpha}$ is $n$-copies distillable using the characterization given by the equation \eqref{AlternativeDefDist}.
\end{proof}

For the particular case $n=1$, the quadratic  form \eqref{Definitionqn} is given by
\begin{equation}
    q^{(1)}(\alpha,C)=\Vert C \Vert_2^2+\alpha \vert \tr C \vert^2,
\end{equation}
 which is positive for every matrix of rank $r$ if  $\alpha \geq -\frac{1}{r}$ , by the inequality \eqref{RelationNormTrace}. 
 In particular for $r=2$, we get the expected boundary value  $\alpha=-\frac{1}{2}$ for the $1$-distillability. This  observation on the rank together with numerics performed on the positivity of these quadratic forms, lead us to the following conjecture on the positivity of the form $q^{(n)}$.

\begin{conjecture}\label{conjecture 2}
    Let $C \in L((\mathbb{C}^d)^{\otimes n})$ be a matrix with $\rk C=r$. Then, $q^{(n)}(\alpha,C)\geq 0$, for every $\alpha \geq -\frac{1}{r}$.
\end{conjecture}
Notice that proving Conjecture \ref{conjecture 2} for $r=1$ and $r=2$, proves Conjecture \ref{conjecture1}, due to  Theorem \ref{Theorem:2distillability}. 
For the rank 1 case, we will show in the next section  that we have positivity of \eqref{Definitionqn} for $\alpha=-1$. To prove that $\alpha=-1$ is indeed the boundary (as established in Conjecture \ref{conjecture 2}),    we now look at what happens to e.g. $q^{(2)}$ for the $2$-distillability, for values $\alpha<-1$. Take $u,v,w \in \mathbb{C}^d$  three normalized vectors with $v \perp w$, and define  the matrix 
    $
        C=\vert u \rangle \langle u \vert \otimes\vert v \rangle \langle w \vert  .
    $    
    Then,
    \begin{equation}\label{rank1counterexample}
        q^{(2)}(-1-\varepsilon,C)=1-(1+\varepsilon)=-\varepsilon.
    \end{equation}

Since the $1$-distillability of Werner states for $\alpha \in ]-1,-\frac{1}{2}[$
implies its $n$-distillability for $n \geq 2$, then by Theorem \ref{Theorem:2distillability} there exists a matrix $C$ with rank 2 such that $q^{(n)}(\alpha,C)<0$. An  explicit example of the saturation of the form $q^{(n)}$ (and hence an alternative proof of the previous statement) is shown in Appendix \ref{appendix1} for $n$ even. For the particular case of the $2$-distillability, using Theorem \ref{Theorem:2distillability}, we can find some Werner states that are not PPT  and $2$-undistillable for any dimension $d\geq 5$.

\begin{corollary}\label{Coro1/4}
 If $\alpha\geq -\frac{1}{2r}$, then
$
q^{(2)}(\alpha,C)\geq 0,$ for every $C \in L(\Hilbert_1 \otimes \Hilbert_2)$ with rank $r$.
As a consequence, $\rho_{\alpha}$ is not $2$-distillable for $\alpha \geq -\frac{1}{4}$.
\end{corollary}
\begin{proof}

For $\alpha \geq 0$ the result is clear, so assume that $\alpha <0$. We bound from below the quadratic form \eqref{Definitionqn} using  inequalities \eqref{Inequality12norm} and \eqref{InequalityPartialTrace1}
\begin{equation}
\begin{split}
q^{(2)}(\alpha,C)& =\Vert C \Vert_{2}^2+ \alpha \left[ \Vert \tr_2 C\Vert_{2}^2+\Vert \tr_1 C \Vert_{2}^2 \right] +\alpha^2 \vert \tr C  \vert^2\\
&\geq \Vert C \Vert_{2}^2+ \alpha \left[ \Vert \tr_2 C\Vert_{1}^2+\Vert \tr_1 C \Vert_{1}^2 \right] +\alpha^2 \vert \tr C \vert^2 \\
 &\geq \Vert C \Vert_{2}^2+ 2\alpha  \Vert C \Vert_{1}^2 +\alpha^2 \vert \tr C \vert^2 \\
 &\geq \left( \frac{1}{r}+2\alpha \right)\Vert C \Vert_1^2+\alpha^2 \vert \tr C \vert^2.
\end{split}
\end{equation}
Thus, if 
$
 \alpha\geq -\frac{1}{2r}.
$
we get $q^{(2)}(\alpha,C)\geq 0$.
\end{proof}

For positive matrices, the a priori boundary value $\alpha =-\frac{1}{r}$  can   actually be improved, since this value does not depend necessarily on the rank (see \cite{Huber} or \cite{Rains}), but as we have seen, this changes for the general case. We will discuss in the next section that for higher ranks, the boundary value for $\alpha$ might not be $\alpha=-\frac{1}{r}$  anymore, since the dimension of the systems also plays an important role.

\section{Partial Trace inequalities}\label{sec:PartialTraceInequalities}

In order to prove the positivity of the quadratic form \eqref{Definitionqn} for rank one matrices for the value $\alpha=-1$, we state the following result, which also shows the underlying structure of these quadratic forms. 

\begin{proposition}\label{Prop:Rank1case}
    For $\Hilbert=\Hilbert_1 \otimes  \Hilbert_2$, the form $q^{(2)}(-1, C)$  is positive for every rank $1$ matrix $C\in L(\Hilbert)$.
\end{proposition}
\begin{proof}
   Let $v,w \in \Hilbert=\Hilbert_1 \otimes \Hilbert_2$,  and write
    \begin{equation*}
        \vert v \rangle \langle w \vert=\sum_{i,j=1}^{n}\vert v_i^1 \rangle \langle w_j^1 \vert \otimes \vert v_i^2 \rangle \langle w_j^2 \vert,
    \end{equation*}
   where $n=\max\{ \dim\Hilbert_1, \dim\Hilbert_2\}$. Note that we can make this assumption by completing the vector with fewer elements with zeros. Now, we compute all the norms
    \begin{subequations}
    \begin{equation}
        \Vert \vert v \rangle \langle w \vert \Vert^2_{2}=\sum_{i,j,k,l=1}^{n} \langle w_j^1,w_l^1 \rangle \langle v_k^1,v_i^1 \rangle \langle w_j^2, w_l^2 \rangle \langle v_k^2,v_i^2 \rangle=\sum_{i,j,k,l=1}^{n} \langle v_k^1 \otimes w_j^1 , v_i^1 \otimes w_l^1 \rangle \langle v_k^2 \otimes w_j^2 , v_i^2 \otimes w_l^2 \rangle 
    \end{equation}
    \begin{equation}
        \Vert \tr_1\vert v \rangle \langle w \vert \Vert^2_{2}=\sum_{i,j,k,l=1}^{n} \langle w_j^1,v_i^1 \rangle \langle v_k^1,w_l^1 \rangle \langle w_j^2, w_l^2 \rangle \langle v_k^2,v_i^2 \rangle=\sum_{i,j,k,l=1}^{n} \langle v_k^1 \otimes w_j^1 , w_l^1 \otimes v_i^1 \rangle \langle v_k^2 \otimes w_j^2 , v_i^2 \otimes w_l^2 \rangle 
    \end{equation}
     \begin{equation}
        \Vert \tr_2 \vert v \rangle \langle w \vert \Vert^2_{2}=\sum_{i,j,k,l=1}^{n} \langle w_j^1,w_l^1 \rangle \langle v_k^1,v_i^1 \rangle \langle w_j^2,v_i^2 \rangle \langle v_k^2,w_l^2 \rangle=\sum_{i,j,k,l=1}^{n} \langle v_k^1 \otimes w_j^1 , v_i^1 \otimes w_l^1 \rangle \langle v_k^2 \otimes w_j^2 , w_l^2 \otimes v_i^2 \rangle 
    \end{equation}
    \begin{equation}
        \left\vert \tr\vert v \rangle \langle w \vert \right\vert^2=\sum_{i,j,k,l=1}^{n} \langle w_j^1, v_i^1 \rangle \langle v_k^1, w_l^1 \rangle \langle w_j^2 , v_i^2 \rangle \langle v_k^2,w_l^2 \rangle=\sum_{i,j,k,l=1}^{n} \langle v_k^1 \otimes w_j^1, w_l^1 \otimes v_i^1 \rangle  \langle v_k^2 \otimes w_j^2, w_l^2 \otimes v_i^2 \rangle ,
    \end{equation}
    \end{subequations}
    and using $(\mathbb{1}-F)^2=2(\mathbb{1}-F)$,
    \begin{subequations}
        \begin{align}
    \frac{1}{4} \left\Vert \sum_{k,j=1}^{n} (v_k^1 \wedge w_j^1) \otimes (v_k^2 \wedge w_j^2) \right\Vert^2&=\\
    & \hspace{-2cm}=\frac{1}{4} \sum_{i,j,k,l=1}^{n} \left\langle v_k^1 \otimes w_j^1, (\mathbb{1}-F)^2(v_i^1 \otimes w_l^1)     \right\rangle  \left\langle  v_k^2 \otimes w_j^2, (\mathbb{1}-F)^2(v_i^2 \otimes w_l^2)     \right\rangle\\
    &\hspace{-2cm}=\sum_{i,j,k,l=1}^{n}\left\langle v_k^1 \otimes w_j^1,v_i^1 \otimes w_l^1- w_l^1 \otimes v_i^1    \right\rangle  \left\langle v_k^2 \otimes w_j^2,v_i^2 \otimes w_l^2- w_l^2 \otimes v_i^2    \right\rangle\\
           &\hspace{-2cm}= \Vert \vert v \rangle \langle w \vert \Vert^2_{2}- \Vert \tr_1\vert v \rangle \langle w \vert \Vert^2_{2}- \Vert \tr_2\vert v \rangle \langle w \vert \Vert^2_{2}+\left\vert \tr\vert v \rangle \langle w \vert \right\vert^2\\
           & \hspace{-2cm}=q^{(2)}(-1,\vert v \rangle \langle w \vert).
        \end{align}
    \end{subequations}

\end{proof}

\begin{remark}\label{remark:AlternativeFormq2}
    Notice that these forms can be written in a shorter way. For example, $ q^{(2)}(-1,\vert v \rangle \langle w \vert)$ can be written as follows:
\begin{equation}
\begin{split}
    q^{(2)}(-1,\vert v \rangle \langle w \vert)&=\frac{1}{4}\langle (\mathbb{1}-F)\otimes(\mathbb{1}-F)F_{23}v \otimes w,(\mathbb{1}-F)\otimes(\mathbb{1}-F)F_{23}v \otimes w \rangle \\ & =\langle v \otimes w,(\mathbb{1}-F_{13})(\mathbb{1}-F_{24}) v \otimes w \rangle,
\end{split}
\end{equation}
where $F_{ij}$ is the operator that flips the components $i$ and $j$ and $F=F_{13}F_{24}$. Similarly the rest.
\end{remark}

\begin{remark}
In a similar way it can be checked that for $\Hilbert=\Hilbert_1 \otimes \hdots \otimes \Hilbert_n$ and
    \begin{equation*}
        C=\vert v \rangle \langle w \vert=\sum_{i,j=1}^{k}\vert v_i^1 \rangle \langle w_j^1 \vert \otimes \hdots \otimes  \vert v_i^n \rangle \langle w_j^n \vert,
    \end{equation*}
    with  $v,w \in \Hilbert$ and $k=\max\{\dim \Hilbert_1, \hdots, \dim \Hilbert_n\}$, then
    \begin{equation}\label{rank1expression}
    q^{(n)}(-1,\vert v \rangle \langle w \vert)=\frac{1}{2^n} \left\Vert \sum_{i,j=1}^k (v_{i}^1 \wedge w_{j}^1) \otimes \hdots \otimes  (v_{i}^n \wedge w_{j}^n) \right\Vert^2 \geq 0.
    \end{equation}
\end{remark}

Changing the antisymmetrizations "$\wedge$" by symmetrizations "$\odot$" in \eqref{rank1expression}, it is possible to generate different rank 1 inequalities. In fact, given $n \geq 2$, there are $2^n$ combinations of symmetrizations and antisymmetrizations,  and each one has associated one quadratic form. These symmetries coincide with the ones introduced  in  \cite{Rains}, and motivate us to introduce the following definition.

\begin{definition}
    Let $\Hilbert=\Hilbert_1 \otimes  \hdots \otimes \Hilbert_n$ and $v\in \{0,1\}^n$, $v=(v_k)_{k=1}^n$, then we define
\begin{equation}\label{Definitionqv}
    q_v(\alpha,C)= \sum_{J \in P(\{1,2, \hdots ,n\})}  \alpha^{\vert J \vert}(-1)^{(\vert J \vert+\sum_{ k \in J} v_k)}\Vert \tr_{J} C \Vert_2^2
\end{equation}
where $P(X)$ is the power set of $X$ and where  we denote $\tr_{\emptyset}=\mathbb{1}$.
\end{definition}
\begin{remark}
    Notice that when $v_i=0$ for $1 \leq i \leq n$, that corresponds with a symmetrization in the tensor factor  $i$ in \eqref{rank1expression}. Conversely, when $v_i=1$, then the tensor factor $i$  in  \eqref{rank1expression} corresponds with an antisymmetrization.
\end{remark}
This definition corresponds with the quadratic form associated to a subfamily of the  universal state inversions, which have been  studied in e.g. \cite{Huber}, \cite{Eltschka} or \cite{Lewenstein2}. Notice that  the vector $v_0=(1,1,  \hdots, 1)$, has the associated quadratic form  $q_{v_0}=q^{(n)}$. For $n=3$, for example, the different classes of forms are
\begin{multline}\label{ExampleIneqR31}
q_{(1,1,1)}(\alpha,C)=\Vert C \Vert_2^2+\alpha(\Vert \tr_1 C \Vert_2^2+\Vert \tr_2 C \Vert_2^2+\Vert \tr_3 C \Vert_2^2)+\\ 
\alpha^2(\Vert \tr_{12} C \Vert_{2}^2+\Vert \tr_{13} C \Vert_2^2+\Vert \tr_{23} C \Vert_2^2)+\alpha^3\vert \tr C \vert^2,
\end{multline}
\begin{multline}\label{ExampleIneqR32}
  q_{(0,1,1)}(\beta,C)=\Vert C \Vert_2^2 +\beta(-\Vert \tr_1 C \Vert_2^2+\Vert \tr_2 C \Vert_2^2+\Vert \tr_3 C \Vert_2^2)+\\ 
  \beta^2(-\Vert \tr_{12} C \Vert_{2}^2-\Vert \tr_{13} C \Vert_2^2+\Vert \tr_{23} C \Vert_2^2)-\beta^3\vert \tr C \vert^2, 
\end{multline}
\begin{multline}\label{ExampleIneqR33}
    q_{(0,0,1)}(\gamma,C)=\Vert C \Vert_2^2 +\gamma(-\Vert \tr_1 C \Vert_2^2-\Vert \tr_2 C \Vert_2^2+\Vert \tr_3 C \Vert_2^2)+\\
    \gamma^2(\Vert \tr_{12} C \Vert_{2}^2-\Vert \tr_{13} C \Vert_2^2-\Vert \tr_{23} C \Vert_2^2)+\gamma^3\vert \tr C \vert^2.
\end{multline}
By choosing the position of the symmetrizations, one can find 2 forms more like \eqref{ExampleIneqR33}, and another 2  more like \eqref{ExampleIneqR32}. At this point we are ready to introduce the main conjecture of this work.

\begin{conjecture}\label{Conjecture 3}
    Let $\Hilbert$ be a finite-dimensional Hilbert space that can be decomposed as $\Hilbert=\Hilbert_1\otimes \hdots \otimes \Hilbert_n$, with $\dim (\Hilbert_i)=d_i \geq 2$. Then, for every $C \in L(\Hilbert)$ with $\rk(C)=r$ and every $v \in \{0,1\}^n$, 
$
    q_v(\alpha,C)\geq 0
$
for \begin{equation}\label{alphaoptimal}
\vert \alpha\vert  \leq \alpha_{opt}=\displaystyle  \frac{1}{\min\{ r, \max\{d_1, \hdots ,d_n\} \} }. 
\end{equation}
\end{conjecture} 

 At this point, we recall that one   of the  quadratic forms  in \eqref{Definitionqv} was originally motivated by the distillability of Werner states.   In the proof of the next result, it can be seen how  the upper bound for $\alpha$ in \eqref{alphaoptimal} in terms of the dimension is connected to the separability of the Werner states.  

\begin{proposition}\label{Prop:DimensionBound}
   For $\vert \alpha\vert \leq \frac{1}{\max\{d_1, \hdots ,d_n\} }$, the Conjecture  \ref{Conjecture 3} holds.
\end{proposition}
\begin{proof}
   Let $\Hilbert=\Hilbert_1 \otimes \hdots \otimes \Hilbert_n$ with $\dim \Hilbert_i=d_i$ and $d=\max_i \{d_i\}$. Let $v \in \{0,1\}^n$ with associated quadratic form $q_v$, so $q_v$ can be written as similarly as we did in the proof of Theorem \ref{Theorem:2distillability}
   \begin{equation}
       q_v(\alpha,C)=\tr[(\mathbb{1}\pm \alpha F_{A_1B_1})\otimes \hdots \otimes (\mathbb{1}\pm \alpha F_{A_nB_n})(C \otimes C^*)\tilde{F}],
   \end{equation}
   with the corresponding choice of signs, and where $\tilde{F}$ is the flip operator in $(\mathbb{C}^{d})^{\otimes n}\otimes (\mathbb{C}^{d})^{\otimes n}$. Now, decompose  as in Theorem \ref{Theorem:2distillability}, $C=\sum_{i=1}^r \vert v_i \rangle \langle w_i \vert$ and obtain again
   \begin{equation}
       (C \otimes C^* \tilde{F})^{\tilde{T}_1}=\vert \psi_C \rangle \langle \psi_C \vert,
   \end{equation}
   where $ \psi_C =\sum_{i=1}^r  v_i  \otimes  w_i $. This allow us to write 
   \begin{equation}
       q_v(\alpha,C)=\langle \psi_C,(\mathbb{1}\pm \alpha F_{A_1B_1})^{T_1}\otimes \hdots \otimes (\mathbb{1}\pm \alpha F_{A_nB_n})^{T_1}\psi_C \rangle.
   \end{equation}
    Since for $\vert \alpha \vert \leq \frac{1}{d}$ the Werner states are separable \cite{Vianna}, in particular they are PPT  and we conclude that $q_v(\alpha,C)\geq 0$ for $\vert \alpha \vert \leq \frac{1}{d}$. Finally, the result holds by considering the embedding of $L(\Hilbert)$ in $L((\mathbb{C}^d)^{\otimes n})$.
\end{proof}

    Notice that for positive matrices, the previous Proposition was already proven in e.g. \cite{Eltschka}, and here present an alternative argument that extends this inequality to general matrices using the separability of Werner states. For the bounds where only the dimension of the Hilbert spaces appear, we can reduce any  quadratic form with any vector to the case of the subvector containing all the $1$'s, i.e. it is   sufficient to prove the result for the vectors of $1$'s only.  For example, for $v=(1,0)$ and $\alpha=-\frac{1}{d}$, $d=\max\{d_1,d_2\}$, the correspondent quadratic form is
    \begin{equation}
       q_{(1,0)}\left(-\frac{1}{d},C\right)=\Vert C \Vert_2^2- \frac{1}{d}\Vert \tr_1 C \Vert_2^2+\frac{1}{d}\Vert \tr_2 C \Vert_2^2-\frac{1}{d^2}\vert \tr C\vert^2.
    \end{equation}
    We modify the quadratic form $q_{(1)}$ and define
    \begin{equation}
           q_{(1)}^{\{\Hilbert_1\}}\left(-\frac{1}{d},C\right)=\Vert C \Vert_2^2- \frac{1}{d}\Vert \tr_1 C \Vert_2^2,
    \end{equation}
    where the upper index in $q$ denotes that we take partial trace on the first system instead of the full trace. This form is positive by \cite{Rastegin} and also
    \begin{equation}
         q_{(1)}^{\{\Hilbert_1\}}\left(-\frac{1}{d},\tr_2 C\right)= \Vert \tr_2 C \Vert_2^2- \frac{1}{d}\vert \tr C \vert^2\geq 0.
    \end{equation}
    As a consequence, we can write
    \begin{equation}\label{decompositionseparable}
        q_{(1,0)}\left(-\frac{1}{d},C\right)=
        q_{(1)}^{\{\Hilbert_1\}}\left(-\frac{1}{d},C\right)+ q_{(1)}^{\{\Hilbert_1\}}\left(-\frac{1}{d},\tr_2 C\right).
    \end{equation}
    However, such a decomposition is no longer valid for the rank, because partial traces do not preserve the rank in general.

Next Theorem shows a version, in terms of partial trace inequalities, of  the progress that we have done in Conjecture \ref{Conjecture 3} for $n=2$. The proof is presented in section \ref{sec:ProofTheorem2}.
\begin{theorem}\label{theo:inequalities}
    Let $C \in L(\Hilbert_1\otimes \Hilbert_2)$, $r=\rk(C)$ and  $d=\max\{d_1,d_2\}$, then the following inequalities hold
    \begin{itemize}
    \item[1.] 
    \begin{equation}\label{DifPartialTraces}
     \left\vert   \Vert \tr_{1} C \Vert_{2}^2- \Vert \tr_{2} C \Vert_{2}^2  \right\vert  \leq \min\{r,d\} \Vert C \Vert_{2}^2-\frac{1}{\min\{r,d\}}\vert \tr(C) \vert^2.
\end{equation}
 \item[2.] 
\begin{equation}\label{Inequality2dim}
    \Vert \tr_{1} C \Vert_{2}^2+ \Vert \tr_{2}C \Vert_{2}^2 \leq d \Vert C \Vert_{2}^2+\frac{1}{d}\vert \tr(C) \vert^2.
\end{equation}
    \end{itemize}
    If, in addition, $C$ can be written as $C=C_1+C_2$ with $\rk(C_1)=1$ and $C_2$ normal such that the vectors spanning the range of $C_1$ and $C_1^*$ are orthogonal to all eigenvectors of $C_2$, then
        \begin{equation}\label{Inequality2distillability}
    \Vert \tr_{1} C \Vert_{2}^2+ \Vert \tr_{2}C \Vert_{2}^2 \leq r \Vert C \Vert_{2}^2+\frac{1}{r}\vert \tr(C) \vert^2.
\end{equation}
\end{theorem}
\begin{remark}
    For the particular case $n=2$, $\rk(C)=2$, the condition $q^{(2)}\left(\alpha=-\frac{1}{2}, C\right)\geq 0$ in  \eqref{Definitionqn} can be rewritten as 
\begin{equation}
    \Vert \tr_1 C \Vert_2^2+  \Vert \tr_2 C \Vert_2^2 \leq 2 \Vert C \Vert_2^2+ \frac{1}{2} \vert \tr C \vert^2,
\end{equation}
so \eqref{Inequality2distillability} is the generalization inequality for a rank $r$ matrix. Since we cannot prove \eqref{Inequality2distillability}  for a general rank 2 matrix yet, the problem of the $2$-distillabilty remains open.
\end{remark}
\begin{remark}\label{rem:PlusMinus}
    Inequality \eqref{DifPartialTraces} for rank $1$ and rank $2$ shows (following the reasoning of Theorem \ref{Theorem:2distillability}) that for every $A,B:  (\mathbb{C}^d)^{\otimes 2} \to \mathbb{C}^2$ and $\psi_C \in \mathbb{C}^2\otimes \mathbb{C}^2$,  the Werner states $\rho_{\alpha}$ satisfy:
    \begin{equation}
        \left\langle \psi_C, (A \otimes B)\left(\rho_{\frac{1}{2}}\otimes \rho_{-\frac{1}{2}}\right)^{T_1}(A \otimes B)^*\psi_C  \right\rangle \geq 0,
    \end{equation}
    and 
\begin{equation}
        \left\langle \psi_C, (A \otimes B)\left(\rho_{-\frac{1}{2}}\otimes \rho_{\frac{1}{2}}\right)^{T_1}(A \otimes B)^*\psi_C  \right\rangle \geq 0,
    \end{equation}
    i.e. $\rho_{\frac{1}{2}}\otimes \rho_{-\frac{1}{2}}$ and $\rho_{-\frac{1}{2}}\otimes \rho_{\frac{1}{2}}$ are $1$-distillable in  $L((\mathbb{C}^d)^{\otimes 2})$. Thus, a positive answer to Conjecture \ref{Conjecture 3} for rank $1$ and  rank $2$ matrices would not  only provide a proof of the distillability of Werner states, but  would also show the distillability properties of tensor product of Werner states.
\end{remark}

\section{3-distillability and tripartite systems inequalities}\label{sec:3Hilbert}

Once we have studied $2$-distillability in depth, in this section we begin a small approach to the $3$-distillability, i.e. the problem of showing the positivity of  $q^{(3)}\left( -\frac{1}{2},C\right)$ for rank $2$ matrices $C$. This problem turns out to be  more challenging as the $2$-distillability case, and for the particular case of a rank 2 matrix we can only  show positivity  for self-adjoint matrices with one positive and one negative eigenvalue.
\begin{proposition}\label{prop:3disti}
    If $C\in L(\Hilbert_1 \otimes \Hilbert_2\otimes \mathcal{H}_3)$ is a self-adjoint  rank $2$ matrix with one positive and one negative eigenvalue, then $q^{(3)}\left( -\frac{1}{2},C\right)\geq 0$.
\end{proposition}
\begin{proof}
Consider the  inversion of a pure state
\begin{multline}
    Q_a^{(3),r}=\vert a \rangle \langle a \vert-\frac{1}{r}\left( \mathbb{1}_{d_1} \otimes \tr_{1}( \vert a \rangle \langle a \vert)+ \mathbb{1}_{d_2} \otimes \tr_{2}(\vert a \rangle \langle a \vert) + \tr_{3}(\vert a \rangle \langle a \vert) \otimes \mathbb{1}_{d_3}\right)+\\
    \frac{1}{r^2}\left( \mathbb{1}_{d_1d_2}\otimes \tr_{1  2}(\vert a \rangle \langle a \vert)+\mathbb{1}_{d_1d_3}\otimes \tr_{13}(\vert a \rangle \langle a \vert)+\tr_{23}(\vert a \rangle \langle a \vert)\otimes \mathbb{1}_{d_2d_3} \right)-\frac{1}{r^3}\Vert a \Vert^2\mathbb{1}_{d_1d_2d_3}.
\end{multline}
for $a \in \Hilbert$. By denoting $\tilde{Q}_a^{(3),r}=P_a^{\perp}Q_a^{(3),r}P_a^{\perp}$, where $P_a^{\perp}$ is again the projection on $\ker(\vert a \rangle \langle a \vert)$, we can upper bound this operator 
\begin{equation}\label{BoundQTripartite}
    \tilde{Q}_a^{(3),r} \leq \frac{1}{r^2}\left(1-\frac{1}{r} \right).
\end{equation}
To prove this, notice that by direct computation
\begin{equation}\label{TracePartition}
    \langle \tr_{I}(\vert a \rangle \langle a \vert ),  \tr_{I}(\vert x \rangle \langle x \vert )\rangle=\Vert \tr_{I^C}(\vert a \rangle \langle x \vert) \Vert_2^2,
\end{equation}
for any partition $I, I^C$ of $\{1, \hdots,n \}$ (in this case $n=3$) so we get 
\begin{subequations}
\begin{align}
    \langle x,-\tilde{Q}_a^{(3),r} x \rangle&=\frac{1}{r^2}q^{(3)}(-1,\vert a \rangle \langle x \vert)-\frac{1}{r^2}\left(1-\frac{1}{r} \right)\Vert a \Vert^2 \Vert  x \Vert^2\\
    &+\frac{1}{r}\left(1-\frac{1}{r} \right)(\Vert \tr_{1 2}(\vert a \rangle \langle x \vert) \Vert^2+\Vert \tr_{1 3}(\vert a \rangle \langle x \vert) \Vert^2+\Vert \tr_{23}(\vert a \rangle \langle x \vert) \Vert^2)\\
    &\geq -\frac{1}{r^2}\left(1-\frac{1}{r} \right)\Vert a \Vert^2 \Vert  x \Vert^2,
\end{align}
since $q^{(3)}(-1,\vert a \rangle \langle x \vert)\geq 0$ by Proposition \ref{Prop:Rank1case}.
\end{subequations}

    Let $C$ be a self-adjoint matrix with spectral decomposition $C=  \vert v_1 \rangle \langle v_1 \vert-\vert v_2\rangle \langle v_2 \vert$, with  $v_1 \perp v_2$, but the vectors are not normalized.
\begin{multline}
    q^{(3)}\left(-\frac{1}{2},C\right)=\sum_{i=1}^2\biggl[ \left(1-\frac{1}{8}\right)\Vert v_i \Vert_2^4-\frac{1}{2}\left( \Vert \tr_1 \vert v_i \rangle \langle v_i \vert \Vert_2^2 +\Vert \tr_2 \vert v_i \rangle \langle v_i \vert \Vert_2^2 +\Vert \tr_3 \vert v_i \rangle \langle v_i \vert \Vert_2^2  \right)\\
    +\frac{1}{4}\left( \Vert \tr_{12} \vert v_i \rangle \langle v_i \vert \Vert_2^2 +\Vert \tr_{13} \vert v_i \rangle \langle v_i \vert \Vert_2^2 +\Vert \tr_{23} \vert v_i \rangle \langle v_i \vert \Vert_2^2  \right) \biggr]
    -2 \langle v_1 ,\tilde{Q}_{v_2}^{(3)} v_1\rangle.
\end{multline}
Finally, using \eqref{BoundQTripartite} with $r=2$ together with \eqref{TracePartition}, we can write
\begin{equation}
\begin{split}
 q^{(3)}\left(-\frac{1}{2},C\right)&=\frac{1}{4}\sum_{i=1}^2\sum_{j=1}^3\left( \Vert v_i \Vert_2^2-\Vert \tr_j \vert v_i \rangle \langle v_i \vert\Vert_2^2 \right)+\frac{1}{8}(\Vert v_1\Vert_2^2-\Vert v_2\Vert_2^2)^2\\
 &\geq 0.
 \end{split}
\end{equation}
\end{proof}

We  also study the other two families of inequalities in tripartite systems \eqref{ExampleIneqR32} and \eqref{ExampleIneqR33} for the case that the parameter depends on the rank of the matrix. Proving the positivity for $\beta,\gamma=-\frac{1}{2}$ for arbitrary rank 1 and rank 2 matrices, would give us similar conclusions to the ones obtained in Remark \ref{rem:PlusMinus}. Here, we will show them only  for  positive semidefinite matrices. We will need  first the following result.
\begin{lemma}\label{Lemma:PartitionPartialTrace}
     Let $C \in L(\Hilbert_1 \otimes \Hilbert_2)$ be a positive semidefinite matrix with $r=\rk(C)$. Then,
      $\displaystyle \Vert \tr_{1} C \Vert_2^2 \geq \frac{1}{r}\Vert \tr_{2} C \Vert_2^2$ 
\end{lemma}
\begin{proof}
  Write the spectral decomposition of $\displaystyle C=\sum_{i=1}^r \vert v_i \rangle \langle v_i \vert$ and by \eqref{TracePartition} we get that 
    \begin{equation}
        \Vert \tr_{1} (\vert v_i \rangle \langle v_i \vert)\Vert_2= \Vert \tr_{2} (\vert v_i \rangle \langle v_i \vert)\Vert_2,
    \end{equation}
         for every $1 \leq i \leq r$. Thus, by \eqref{TracePartition} again we can express everything in terms of the partial trace over the second system
         \begin{multline}
         \Vert \tr_{1} C \Vert_2^2 -\frac{1}{r}\Vert \tr_{2} C \Vert_2^2 \geq \\ \left(1-\frac{1}{r}\right)\sum_{i=1}^r\Vert \tr_{2} (\vert v_i \rangle \langle v_i \vert)\Vert_2^2+2\sum_{\substack{i,j=1\\ i>j}}^r\Vert \tr_{2} (\vert v_i \rangle \langle v_j \vert)\Vert_2^2-\frac{2}{r}\sum_{\substack{i,j=1\\ i>j}}^r \Vert \tr_{2} (\vert v_i \rangle \langle v_i \vert)\Vert_2 \Vert \tr_{2} (\vert v_j \rangle \langle v_j \vert)\Vert_2,
     \end{multline}
     To conclude, define the polynomial 
\begin{equation}\label{PolynomialPositivity}
    p_{r}(x)=(r-1)\sum_{i=1}^{r} x_i^2-2\sum_{\substack{i,j=1 \\ i > j}}^{d_1}x_ix_j =\sum_{\substack{i,j=1 \\i >j }}^r(x_i-x_j)^2\geq 0,
\end{equation}
so we get
\begin{subequations}
\begin{align}
    \Vert \tr_{1} C \Vert_2^2 -\frac{1}{r}\Vert \tr_{2} C \Vert_2^2 &\geq \frac{1}{r}p_r\left( \Vert \tr_{2} (\vert v_1 \rangle \langle v_1 \vert)\Vert_2, \hdots, \Vert \tr_{2} (\vert v_r \rangle \langle v_r \vert) \Vert_2\right)+2\sum_{\substack{i,j=1\\ i>j}}^r\Vert \tr_{2} (\vert v_i \rangle \langle v_j \vert)\Vert_2^2\\&\geq 0.
\end{align}
\end{subequations}
\end{proof}

\begin{proposition}\label{prop:IneqR32}
    The conjecture over the class of  forms of  \eqref{ExampleIneqR32} and \eqref{ExampleIneqR33} for $\beta,\gamma=-\frac{1}{r}$,  holds for positive matrices.
\end{proposition}
\begin{proof}
 
 We will start with the proof of the family of inequalities \eqref{ExampleIneqR32}.
 We will just prove \eqref{ExampleIneqR32}, the other 2 are analogous. In this case the associated linear operator is
\begin{multline}
    P_a^{(3),r}=\vert a \rangle \langle a \vert+\frac{1}{r}\left( \mathbb{1}_{d_1} \otimes \tr_{1}( \vert a \rangle \langle a \vert)- \mathbb{1}_{d_2} \otimes \tr_{2}(\vert a \rangle \langle a \vert) - \tr_{3}(\vert a \rangle \langle a \vert) \otimes \mathbb{1}_{d_3} \right)+\\
    \frac{1}{r^2}\left( -\mathbb{1}_{d_1d_2}\otimes \tr_{1 2}(\vert a \rangle \langle a \vert)-\mathbb{1}_{d_1d_3}\otimes \tr_{1 3}(\vert a \rangle \langle a \vert)+\tr_{2 3}(\vert a \rangle \langle a \vert)\otimes \mathbb{1}_{d_2 d_3} \right)+\frac{1}{r^3}\Vert a \Vert^2\mathbb{1}_{d_1d_2d_3}.
\end{multline}
This operator is bounded from below by
\begin{equation}
    P_a^{(3),r} \geq \frac{1-r^2}{r^3}\Vert a \Vert^2,
\end{equation}
on $\ker(\vert a \rangle \langle a \vert)$ by computing the expectation value 
\begin{multline}
    \langle x, P_a^{(3),r} x \rangle=\frac{1}{r^2}p_{-1}(\vert a \rangle \langle x \vert)+\frac{1-r^2}{r^3}\Vert a \Vert^2 \Vert x \Vert^2+ \left( 
 1-\frac{1}{r^2}\right) \vert \langle a ,x \rangle \vert^2 \\ +\frac{1}{r}\left(1-\frac{1}{r}\right)(\Vert a \Vert^2 \Vert x \Vert^2-\Vert \tr_{1 2}(\vert a \rangle \langle x \vert)\Vert^2- \Vert \tr_{13}(\vert a \rangle \langle x \vert)\Vert^2+\Vert \tr_{2 3}(\vert a \rangle \langle x \vert)\Vert^2)\geq 0,
\end{multline}
since
\begin{multline}
        \Vert a \Vert^2 \Vert x \Vert^2-\Vert \tr_{1  2}(\vert a \rangle \langle x \vert)\Vert^2- \Vert \tr_{1  3}(\vert a \rangle \langle x \vert)\Vert^2+\Vert \tr_{23}(\vert a \rangle \langle x \vert)\Vert^2
       = \\ \hspace{-3.2cm}= \langle a \otimes x,(1-F_{14}F_{25})(1-F_{14}F_{36})a \otimes x \rangle \geq 0.
\end{multline}

     Let $C$  be a positive matrix with rank $r$ and write its (non-normalized) spectral decomposition $C=\sum_{i=1}^r \vert v_i \rangle \langle v_i \vert $, then using \eqref{TracePartition} we can write
    \begin{multline}\label{Coro:Form3p}
        q_{(0,1,1)}^{(3)}\left(-\frac{1}{r},C\right)=\left(1-\frac{1}{r}-\frac{1}{r^2}+\frac{1}{r^3}  \right)\sum_{i=1}^r \Vert v_i \Vert^4 +2\sum_{\substack{i,j=1\\ i >j}}^r\langle v_j,P_{v_i}^{(3),r}v_j \rangle    \\ 
        +\frac{1}{r}\left(1+ \frac{1}{r}\right)\sum_{i=1}^r\left( \Vert v_i \Vert^4-\Vert \tr_{1}(\vert v_i \rangle \langle v_i \vert)\Vert^2 -\Vert \tr_{2}(\vert v_i \rangle \langle v_i \vert)\Vert^2+\Vert \tr_{3}(\vert v_i \rangle \langle v_i \vert)\Vert^2  \right).
    \end{multline}       
    To conclude, the first to terms in \eqref{Coro:Form3p} are positive since they are lower bounded by the expression
    \begin{equation}
        \frac{r^2-1}{r^3}\left( (r-1)\sum_{i=1}^r \Vert v_i \Vert^4-2  \sum_{\substack{i,j=1\\ i >j}}^r \Vert v_i \Vert^2\Vert v_j \Vert^2    \right) \geq 0,
    \end{equation}
    and the last one is equal to
    \begin{equation}
        \frac{1}{r}\left(1+ \frac{1}{r}\right)\sum_{i=1}^r\langle v_i \otimes v_i,(1-F_{14})(1-F_{25})(1+F_{36}) v_i \otimes v_i \rangle \geq 0,
    \end{equation}
    which concludes the first part. For \eqref{ExampleIneqR33}, the proof follows then by Lemma \ref{Lemma:PartitionPartialTrace} writing 
\begin{multline}
     q_{(0,0,1)}\left(-\frac{1}{r},C\right)=\left(\Vert C \Vert_2^2-\frac{1}{r}\Vert \tr_{3} C \Vert_2^2 +\frac{1}{r^2}\Vert \tr_{1  2} C \Vert_2^2 -\frac{1}{r^3}\vert \tr C \vert^2 \right)\\ +\frac{1}{r}\left(\Vert \tr_{1} C \Vert_2^2-\frac{1}{r}\Vert \tr_{2 3} C \Vert_2^2 \right)+\frac{1}{r}\left( \Vert \tr_{2} C \Vert_2^2- \frac{1}{r}\Vert \tr_{1  3} C \Vert_2^2 \right) \geq 0.
\end{multline}

\end{proof}

\begin{remark}
    Proposition \ref{prop:IneqR32} shows that the positivity of the quadratic forms associated to the state inversions might also depend on the rank and not only on the dimension. Similar considerations might also hold for the state inversion maps.
\end{remark}

\section{Inequalities for p-Schatten norms}\label{sec:NumericalResults}
Finally, in this last section, we set a more general function than \eqref{Definitionqv}, which will also depend on the norm and the exponent as follows:
\begin{definition}
    Let $\Hilbert=\Hilbert_1 \otimes  \hdots \otimes \Hilbert_n$ and $v\in \{0,1\}^n$. Define for $p\geq 1$, $\gamma \geq 1$, $\alpha \in \mathbb{R}$, $C \in L(\Hilbert)$ and $v \in \{0,1\}^n$ the map
\begin{equation}\label{Definitionqvgeneral}
    q_v(p,\gamma,\alpha,C)=   \sum_{J \in P(\{1,2, \hdots ,n\})  }\alpha^{\vert J \vert}(-1)^{(\vert J \vert+\sum_{ k \in J} v_k)}\Vert \tr_{J} C \Vert_p^{\gamma}.
\end{equation}
\end{definition}
The objective is to study wether it is possible to  extend  Conjecture \ref{Conjecture 3}  to inequalities depending on $p$, $\gamma$, $r=\rk(C)$ or  $d=\max\{d_1, \hdots, d_n\}$ ,  instead of only depending on the rank $r$ and $d$, i.e, we want to study if it is possible to introduce a function $\alpha_v(p,\gamma,r,d)$ that provides tight bounds for the positivity of \eqref{Definitionqvgeneral}. For the particular case of $p=2$ and $\gamma=2$, $ \alpha_v(2,2,r,d)=\alpha_{opt}$ given by \eqref{alphaoptimal}.

For $v=(1)$, the bound with the dimension was actually studied in \cite{Rastegin}, where it was  proved that
\begin{equation}\label{Raste}
    \Vert \tr_i C \Vert_p \leq d^{\frac{p-1}{p}} \Vert C \Vert_p,
\end{equation}
and  is also possible to obtain  a rank bound   using the Hölder inequality for the Schatten $p$-norms
\begin{equation}
    \Vert \tr_i C \Vert_p\leq \Vert \tr_i C \Vert_1 \leq \Vert C \Vert_1 \leq \Vert \mathbb{1}_r \Vert_{p'}\Vert C \Vert_p =r^{\frac{p-1}{p}}\Vert C \Vert_p,
\end{equation}
where $p'$ is the dual Hölder index of $p$ satisfying $\frac{1}{p}+\frac{1}{p'}=1$.  Thus, 
\begin{equation}
    \alpha_{(1)}(p,\gamma,r,d)=\frac{1}{\min\left\{ r^{\gamma \frac{p-1}{p}},d^{\gamma \frac{p-1}{p}} \right\}}
\end{equation}
is a tight bound satisfying
\begin{equation}
    \Vert C \Vert_p^{\gamma}-\alpha_{(1)}(p,\gamma,r,d)\vert \tr C \vert^{\gamma}\geq 0.
\end{equation}

For the vector $v=(1,1)$ it was proved in \cite{Audenaert} that for  any state $\rho$, $p>1$ and  $\gamma=1,p$ 
\begin{equation}
    \Vert \rho \Vert_p^{\gamma}-\Vert \tr_1 \rho \Vert_p^{\gamma}-\Vert \tr_2 \rho \Vert_p^{\gamma} +\vert \tr \rho \vert^{\gamma} \geq 0,
\end{equation}
i.e. $\alpha_{(1,1)}(p,\gamma,r,d)=1$, for $\gamma=1,p$ when restricted to states. This was used to show the subadditivity of the Tsallis entropy. However, for fixed dimensions and different values of $p$ and $\gamma$ and a general  full-rank matrix $C$, the next graphic shows the evolution of the quantity $\alpha_v(p,\gamma)$ which is the quantity  analogous to $\alpha_{opt}$ introduced in \eqref{alphaoptimal} but now dependent on $p$ and $\gamma$ when $r$ and $d$ are fixed, in the systems $\mathbb{R}^2\otimes \mathbb{R}^2$ and $\mathbb{R}^2\otimes \mathbb{R}^3$, which seems to have continuous dependence with respect to  $p$ and $\gamma$.
\begin{figure}[H]
    \centering
    \begin{subfigure}{0.49\textwidth}
    \includegraphics[scale=0.4]{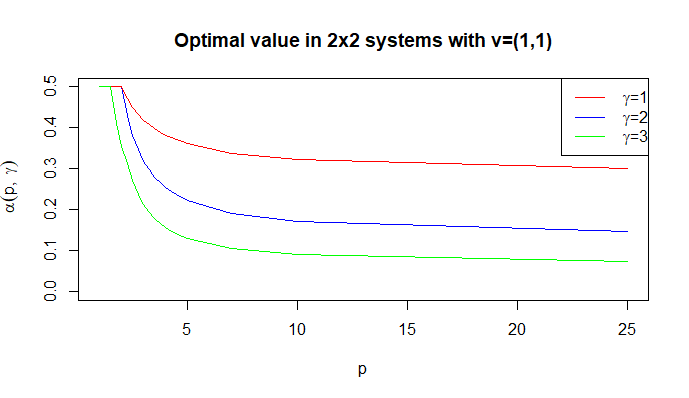}
    \end{subfigure}
        \begin{subfigure}{0.49\textwidth}
    \includegraphics[scale=0.4]{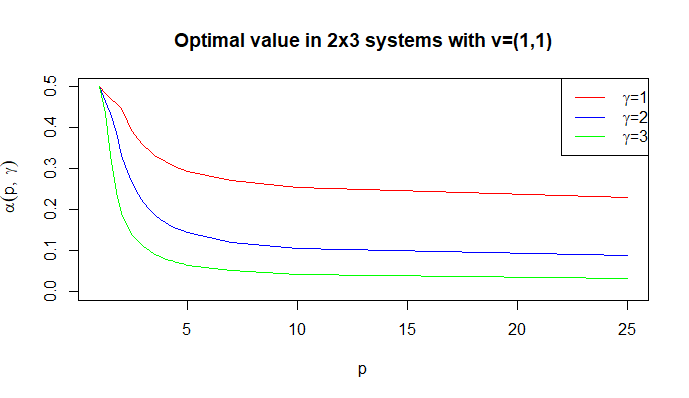}
    \end{subfigure}
    \caption{Optimal values for $v=(1,1)$ for different values of $p$ and $\gamma$ in $\mathbb{R}^2\otimes \mathbb{R}^2$ and $\mathbb{R}^2\otimes \mathbb{R}^3$ }
    \label{fig:plot}
\end{figure}
This shows that there are large families of partial trace inequalities that remain to be studied that generalize the ones of Conjecture \ref{Conjecture 3}, when we also take into account the $p$-norms and the exponents $\gamma$.

\section{Proof of Theorem \ref{theo:inequalities}}\label{sec:ProofTheorem2}

First of all, by  Proposition \ref{Prop:DimensionBound}, 
 the inequalities \eqref{Inequality2dim} and the bounds with the dimension of the Hilbert spaces of \eqref{DifPartialTraces} are proved, so only the bounds with the ranks have to be proved. We will divide the proof into two parts: first, we will prove \eqref{DifPartialTraces} and then \eqref{Inequality2distillability}. Before proving \eqref{DifPartialTraces}, we will need an auxiliary result that allow us to bound the difference of partial traces in a tight way. For this purpose we will make use of the creation and annihilation operators introduced in \eqref{def:creation}, \eqref{def:annihilation}.

\begin{proposition}\label{theo:PartialTraceCraeationAnnihilation}
    Let $v,w \in \Hilbert=\Hilbert_1 \otimes \Hilbert_2$, $d_1=\dim \Hilbert_1$, $d_2=\dim \Hilbert_2$, then
    \begin{equation}
        \mathbb{1}_{d_1} \otimes \tr_1(\vert v \rangle \langle w \vert)=\frac{1}{2}[a_+(w)F_{24}a_+^*(v)+a_-(w)F_{24}a_-^*(v)]\vert_{\Hilbert},
    \end{equation}
    \begin{equation}
        \tr_2(\vert v \rangle \langle w \vert)\otimes \mathbb{1}_{d_2}=\frac{1}{2}[a_+(w)F_{24}a_+^*(v)-a_-(w)F_{24}a_-^*(v)]\vert_{\Hilbert},
    \end{equation}
where    $F_{24}$ is the flip operator that exchanges components 2 and 4.
\end{proposition}
\begin{proof}
Without loss of generality, we can assume that the vectors are normalized. In order to simplify this proof, we will use here the bra-ket notation understanding $\vert v \rangle$ as a vector (not necessarily normalized) with associated functional $\langle v \vert$.    Write $\vert v\rangle=\displaystyle \sum_{i=1}^{n} \vert  v_i^1 \rangle \vert v_i^2 \rangle$ and $\vert w \rangle=\displaystyle \sum_{j=1}^{n} \vert w_j^1 \rangle \vert  w_j^2\rangle$, where we can assume again that $n$ is the same. We will prove 
    \begin{equation}
         \mathbb{1}_{d_1} \otimes \tr_1(\vert v \rangle \langle w \vert)- \tr_2(\vert v \rangle \langle w \vert)\otimes \mathbb{1}_{d_2}=a_-(\langle w\vert )F_{24}a_-^*(\vert v\rangle)\vert_{\Hilbert},
    \end{equation}
    and 
    \begin{equation}\label{Theo:BosonicPartialTrace}
         \mathbb{1}_{d_1} \otimes \tr_1(\vert v \rangle \langle w \vert)+ \tr_2(\vert v \rangle \langle w \vert)\otimes \mathbb{1}_{d_2}=a_+(\langle w\vert )F_{24}a_+^*(\vert v\rangle)\vert_{\Hilbert}.
    \end{equation}
    For the first one, let $\displaystyle \vert x\rangle=\sum_{k=1}^{n} \vert x_k^1 \rangle \vert x_k^2 \rangle$, then
    \begin{subequations}
    \begin{align}
         \left[\mathbb{1}_{d_1} \otimes \tr_1(\vert v \rangle \langle w \vert)- \tr_2(\vert v \rangle \langle w \vert)\otimes \mathbb{1}_{d_2} \right]\vert x \rangle&=\\
         &\hspace{-3cm}=\sum_{i,j,k=1}^{n} \langle w_j^1, v_i^1 \rangle \langle w_j^2,x_k^2 \rangle \vert x_k^1 \rangle\vert v_i^2\rangle -\langle w_j^2, v_i^2 \rangle \langle w_j^1,x_k^1 \rangle \vert v_i^1\rangle \vert x_k^2\rangle \\
        &\hspace{-3cm}=\sum_{i,j,k=1}^{n} \left(\langle w_j^1\vert \langle w_j^2\vert  \otimes \mathbb{1}_{d_1}\otimes \mathbb{1}_{d_2} \right)\left( \vert v_i^1 \rangle \vert x_k^2 \rangle  \vert x_k^1 \rangle\vert v_i^2\rangle- \vert x_k^1 \rangle \vert v_i^2 \rangle \vert v_i^1\rangle \vert  x_k^2\rangle   \right)\\
        &\hspace{-3cm}=\left(\langle w \vert \otimes \mathbb{1}_{d_1}\otimes \mathbb{1}_{d_2} \right)  F_{24}(\vert v \rangle \vert x\rangle- \vert x \rangle \vert v\rangle)\\
        &\hspace{-3cm}=\left(\langle w \vert \otimes \mathbb{1}_{d_1}\otimes \mathbb{1}_{d_2} \right) F_{24}(\mathbb{1}-F)(\vert v \rangle  \vert  x \rangle),\\ 
    \end{align}
    \end{subequations}
    where $F=F_{13}F_{24}$. To conclude the proof, we use that $(1-F)^2=2(1-F)$ together with the fact that $[F_{24},F]=0$, and we can write 
\begin{subequations}
    \begin{align}
         \left[\mathbb{1}_{d_1} \otimes \tr_1(\vert v \rangle \langle w \vert)- \tr_2(\vert v \rangle \langle w \vert)\otimes \mathbb{1}_{d_2} \right]\vert x \rangle&=\\
         &\hspace{-3cm}=2\left(\langle w \vert \otimes \mathbb{1}_{d_1}\otimes \mathbb{1}_{d_2} \right) P_-F_{24}P_-(\vert v \rangle  \vert  x \rangle)\\
         &\hspace{-3cm}=a_-(\langle w\vert )F_{24}a_-^*(\vert v\rangle)\vert x\rangle,
    \end{align}
\end{subequations}
    using the definition of fermionic creation and annihilation operators restricted to one copy of the space given in \eqref{def:creation} and \eqref{def:annihilation}. The inequality \eqref{Theo:BosonicPartialTrace} is analogous.
\end{proof}

Due to linearity, this result can be extended to any $C \in L(\Hilbert)$, resulting in the operator $\mathbb{1}_{d_1} \otimes \tr_{1}(C)-\tr_{2}(C)\otimes \mathbb{1}_{d_2}$ having a "fermionic character", while that the operator $\mathbb{1}_{d_1} \otimes  \tr_{1}(C)+\tr_{2}(C)\otimes \mathbb{1}_{d_2}$ has a "bosonic character". From the fermionic one, we can obtain the following result:
\begin{corollary}\label{Coro:PartialTracesNorm}
For any matrix $C \in L(\Hilbert_1 \otimes \Hilbert_2)$ with rank $r$,
\begin{equation}
    \Vert \mathbb{1}_{d_1} \otimes \tr_{1}(C)-\tr_{2}(C)\otimes \mathbb{1}_{d_2} \Vert_{\infty}\leq \sum_{i=1}^r \sigma_i=\Vert C \Vert_1,
\end{equation}
where $\{\sigma_i\}_{i=1}^r$ is the set of singular values of $C$.
    In particular, for $v,w \in \Hilbert$, then \begin{equation}
        \Vert \mathbb{1}_{d_1} \otimes \tr_{1}(\vert v \rangle \langle w \vert)-\tr_{2}(\vert v \rangle \langle w \vert)\otimes \mathbb{1}_{d_2} \Vert_{\infty}\leq \Vert v \Vert \Vert w \Vert.
    \end{equation}
\end{corollary}

This result follows from the previous Proposition, the singular value decomposition, and the fact that $\Vert a_-(v)\Vert_{\infty} =\Vert a_-^*(v)\Vert_{\infty} =\Vert v \Vert$ (see \cite{Bratteli}).

\begin{proof}[Proof of inequality \eqref{DifPartialTraces}]
   To show inequality \eqref{DifPartialTraces}, we need to show that both $q_{(1,0)}\left(-\frac{1}{r},C\right),q_{(0,1)}\left(-\frac{1}{r},C\right)\geq 0$. We will only show the first, since the other is analogous, i.e. we will prove
    \begin{equation}
        q_{(1,0)}\left(-\frac{1}{r},C\right)=\Vert C \Vert_2^2-\frac{1}{r}\Vert \tr_1 C \Vert_2^2+\frac{1}{r}\Vert \tr_2 C \Vert_2^2-\frac{1}{r^2}\vert \tr C \vert^2\geq 0.
    \end{equation}
   Using the singular value decomposition of $C$, we can write
    \begin{equation}
        C=\sum_{i=1}^r \vert v_i \rangle \langle w_i  \vert,
    \end{equation}
    where $\{v_i\}_{i=1}^r$ and $\{w_i\}_{i=1}^r$ are orthogonal systems of $\Hilbert=\mathcal{H}_1 \otimes \mathcal{H}_2$ (note that the vectors are not normalized, since they absorb the singular value). Write for every $1 \leq i \leq r$,
    \begin{equation}
        q_{(1,0)}\left(-\frac{1}{r},\vert v_i \rangle \langle w_i \vert\right)=\frac{1}{r} q_{(1,0)}\left(-1,\vert v_i \rangle \langle w_i  \vert \right)+\left(1-\frac{1}{r} \right)\Vert v_i \Vert^2 \Vert w_i \Vert^2+\frac{r-1}{r^2} \vert \langle v_i,w_i \rangle \vert^2,
    \end{equation}
    where 
\begin{equation}
    q_{(1,0)}\left(-1,\vert v_i \rangle \langle w_i \vert\right)=\langle v_i \otimes w_i,(\mathbb{1}-F_{13})(\mathbb{1}+F_{24}) v_i \otimes w_i \rangle\geq 0,
\end{equation}
analogously to Remark \ref{remark:AlternativeFormq2}. Then,
    \begin{subequations}
    \begin{equation}\label{RealPartFormp2}
        q_{(1,0)}\left(-\frac{1}{r},C\right)=\frac{1}{r}\sum_{i=1}^r q_{(1,0)}\left(-1,\vert v_i \rangle \langle w_i  \vert\right)+\left(1-\frac{1}{r} \right)\sum_{i=1}^r\Vert v_i \Vert^2 \Vert w_i \Vert^2+\frac{r-1}{r^2}\sum_{i=1}^r \vert \langle v_i,w_i \rangle \vert^2
    \end{equation}
    \begin{equation}\label{RealPartFormp}
        +\frac{2}{r}Re\sum_{\substack{i,j=1\\ i>j}}^r \left[ -\langle \tr_{1}(\vert v_i \rangle \langle w_i  \vert),\tr_{1}(\vert v_j \rangle \langle w_j  \vert)\rangle+
        \langle \tr_{2}(\vert v_i \rangle \langle w_i  \vert),\tr_{2}(\vert v_j \rangle \langle w_j  \vert) \rangle  -\frac{1}{r}\langle v_i,w_i \rangle \langle w_j,v_j \rangle \right].
    \end{equation}
    \end{subequations}
    Now, the key idea is to use the norms and the absolute value of the inner products in \eqref{RealPartFormp2} to bound the term \eqref{RealPartFormp}. We proceed as follows:
    \begin{subequations}
    \begin{align}
         \frac{2}{r}\biggl\vert Re\sum_{\substack{i,j=1\\ i>j}}^r \left[ \langle -\tr_{1}(\vert v_i \rangle \langle w_i  \vert),\tr_{1}(\vert v_j \rangle \langle w_j  \vert) \rangle +
        \langle \tr_{2}(\vert v_i \rangle \langle w_i  \vert),\tr_{2}(\vert v_j \rangle \langle w_j  \vert) \rangle \right] \biggr\vert&= \\
        & \hspace{-7cm}\leq \frac{2}{r}\sum_{\substack{i,j=1\\ i>j}}^r \left\vert \langle \vert v_i \rangle \langle w_i  \vert,- \mathbb{1}_{d_1} \otimes \tr_{1}(\vert v_j \rangle \langle w_j  \vert)+\tr_{2}(\vert v_j \rangle \langle w_j  \vert)\otimes \mathbb{1}_{d_2} \rangle \right\vert \\
        &\hspace{-7cm}\leq \frac{2}{r}\sum_{\substack{i,j=1\\ i>j}}^r \Vert  v_i \Vert \Vert w_i \Vert  \Vert \mathbb{1}_{d_1} \otimes \tr_{1}(\vert v_j \rangle \langle w_j  \vert)-\tr_{2}(\vert v_j \rangle \langle w_j  \vert)\otimes \mathbb{1}_{d_2} \Vert_{\infty}\\
         & \hspace{-7cm}\leq\frac{2}{r}\sum_{\substack{i,j=1\\ i>j}}^r \Vert v_i \Vert \Vert w_i \Vert \Vert v_j \Vert \Vert w_j \Vert,
    \end{align}
    \end{subequations}
where we used Corollary \ref{Coro:PartialTracesNorm}. Finally, considering again the polynomial \eqref{PolynomialPositivity} given by
\begin{equation*}
    p_{r}(x)=(r-1)\sum_{i=1}^{r} x_i^2-2\sum_{\substack{i,j=1 \\ i > j}}^{d_1}x_ix_j =\sum_{\substack{i,j=1 \\i >j }}^r(x_i-x_j)^2\geq 0,
\end{equation*}
we obtain
\begin{multline}
    q_{(1,0)}\left(-\frac{1}{r},C\right)\geq \\ \geq \frac{1}{r}\sum_{i=1}^r q_{(1,0)}\left(-1,\vert v_i \rangle \langle w_i  \vert\right)+\frac{1}{r}p_r(\Vert v_1\Vert \Vert w_1 \Vert, \hdots,\Vert v_r\Vert \Vert w_r \Vert )+\frac{1}{r^2}p_r(\vert \langle v_1,w_1 \rangle \vert, \hdots ,\vert \langle v_r,w_r \rangle \vert),
\end{multline}
which is positive, so the result follows.
\end{proof}

In order to show  \eqref{Inequality2distillability}, one could  think of using the bosonic creation and annihilation operators as we did in the first part. However, this technique does not seem to  work, since these are bounded by the square root of the number operator (\cite{Bratteli}), but we will prove it for matrices of the form sum of a rank 1 plus a normal matrix making use of a different strategy.  Before going into the proof of \eqref{Inequality2distillability}, consider first the following inversion of a pure state
\begin{equation}\label{OpeartorQar}
         Q^r_a=\vert a \rangle \langle a \vert-\frac{1}{r}\left(\mathbb{1}_{d_1} \otimes \tr_{1}(\vert a \rangle \langle a \vert)+\tr_{2}(\vert a \rangle \langle a \vert)\otimes \mathbb{1}_{d_2}\right)+\frac{1}{r^2}\tr(\vert a \rangle \langle a \vert)\mathbb{1}_{d_1d_2},
\end{equation}
which is self-adjoint, for $a \in \Hilbert$. We make use again of \eqref{TracePartition}. i.e.
\begin{equation*}
    \langle \tr_{1}(\vert a \rangle \langle a \vert ),  \tr_{1}(\vert x \rangle \langle x \vert )\rangle=\Vert \tr_{2}(\vert a \rangle \langle x \vert) \Vert_2^2,
\end{equation*}
for every $x \in \Hilbert$. We can obtain then a  bound for the spectral radius on $\ker(\vert a \rangle \langle a \vert)$ as follows: Let $x \in \ker(\vert a \rangle \langle a \vert)$, then
\begin{align}
     \langle x, Q^r_a x \rangle &= 
     \frac{1}{r^2}\Vert a \Vert^2\Vert x \Vert^2  + \vert \langle a ,x \rangle \vert^2-\frac{1}{r}\Vert \tr_{1}(\vert a \rangle \langle x \vert) \Vert_{2}^2-\frac{1}{r}\Vert \tr_{2}(\vert a \rangle \langle x \vert) \Vert_{2}^2\\ 
     &=\frac{1}{r}q^{(2)}(-1,\vert a \rangle \langle x \vert)-\frac{1}{r}\left(1-\frac{1}{r}\right)\Vert a \Vert^2\Vert x \Vert^2,  \label{ExpectedValueQa}
    \end{align}
and conversely
\begin{equation}
    \left\langle x, \left(\frac{1}{r^2} \Vert a \Vert^2-Q^r_a \right)x \right\rangle=\frac{1}{r} \Vert \tr_{1}(\vert a \rangle \langle x \vert) \Vert_{2}^2+\frac{1}{r}\Vert \tr_{2}(\vert a \rangle \langle x \vert) \Vert_{2}^2\geq 0,
\end{equation}
    so
    \begin{equation}\label{BoundQa}
        -\frac{1}{r}\left(1-\frac{1}{r}\right)\Vert a \Vert^2\leq Q_a^r \leq \frac{1}{r^2}\Vert a \Vert^2
    \end{equation}
on $\ker(\vert a \rangle \langle a \vert)$. In particular, if we denote $P_{a^{\perp}}$ the projection onto $\ker(\vert a \rangle \langle a \vert)$, then
\begin{equation}\label{operatornormq2}
     \tilde{Q}_a^r=P_{a^{\perp}} Q_a^r P_{a^{\perp}}
\end{equation}
is self-adjoint and $\Vert  \tilde{Q}_a^r \Vert_{\infty} \leq \frac{1}{r}\left(1-\frac{1}{r}\right)\Vert a \Vert^2$, for $r\geq 2$.

\begin{proof}[Proof of inequality \eqref{Inequality2distillability}]
 The case $r=1$ was proven in Proposition \ref{Prop:Rank1case}, so we can assume that $r\geq 2$. Suppose that $C=C_1+C_2$ with $C_1=\vert v_1 \rangle \langle w_1 \vert$ such that both $v_1,w_1$ are orthogonal to all eigenvectors of the normal matrix $C_2$. Using the spectral decomposition in $C_2$ we can write
\begin{equation} 
    C=\vert v_1 \rangle \langle w_1 \vert+\sum_{i=2}^r \varepsilon_i \vert v_i \rangle \langle v_i \vert,
    \end{equation}
    where $\varepsilon_i \in \mathbb{C}$, and we can assume that $\vert \varepsilon_i \vert =1$ for every $i$,  and the vectors $v_i$ and $w_1$ are not normalized for $1\leq i \leq r$. Our objective is to show that for this choice of $C$,
    \begin{equation}
        q_{(1,1)}\left(-\frac{1}{r},C\right)=q^{(n)}\left(-\frac{1}{r},C\right)  =\Vert C \Vert_2^2-\frac{1}{r}\Vert \tr_1 C \Vert_2^2-\frac{1}{r}\Vert \tr_2 C \Vert_2^2+\frac{1}{r^2}\vert \tr C \vert^2 \geq 0.
    \end{equation}
    Similarly as we did in the proof of inequality \eqref{DifPartialTraces}, we write the quadratic form acting on each rank one matrix
    \begin{equation}
       q_{(1,1)} \left(-\frac{1}{r},\vert v_1 \rangle \langle w_1 \vert \right)=\frac{1}{r}q_{(1,1)} \left(-1,\vert v_1 \rangle \langle w_1 \vert \right)+\left( 1-\frac{1}{r} \right)\Vert v_1 \Vert^2 \Vert w_1 \Vert^2 -\frac{r-1}{r^2}\vert \langle v_1 ,w_1 \rangle  \vert^2, 
    \end{equation}
        \begin{equation}
       q_{(1,1)} \left(-\frac{1}{r},\vert v_i \rangle \langle v_i \vert \right)=\frac{1}{r}q_{(1,1)} \left(-1,\vert v_i \rangle \langle w_i \vert \right)+\left( 1-\frac{1}{r} \right)^2\Vert v_i \Vert^4,
    \end{equation}
    for every $2 \leq i \leq r$, so we obtain
    \begin{multline}
        q_{(1,1)}\left(-\frac{1}{r},C\right) =\frac{1}{r}\left[q_{(1,1)}(-1,\vert v_1 \rangle \langle w_1 \vert)+\sum_{i=2}^r q_{(1,1)}(-1,\vert v_i \rangle \langle v_i \vert)  \right]+ \left(1-\frac{1}{r}\right)\Vert v_1 \Vert^2 \Vert w_1 \Vert^2\\
-\frac{r-1}{r^2}\vert \langle v_1 ,w_1 \rangle  \vert^2 +\left(1-\frac{1}{r}\right)^2\sum_{i=2}^r \Vert v_i \Vert^4
        +2 \sum_{i=2}^r Re\left[\varepsilon_i \langle v_1,Q_{v_i}^r w_1 \rangle \right]+2 \sum_{\substack{i,j=2\\ i>j}}^rRe\left[\varepsilon_i \overline{\varepsilon_j} \langle v_j,Q_{v_i}^r v_j \rangle \right], 
        \end{multline}
where $q_{(1,1)}(-1,\vert v_1 \rangle \langle w_1 \vert),q_{(1,1)}(-1,\vert v_i \rangle \langle v_i \vert)\geq 0$ due to Proposition \ref{Prop:Rank1case} (or due to Remark \ref{remark:AlternativeFormq2}) for every $2 \leq i \leq r$.    Use now the bound of the operator \eqref{operatornormq2}
\begin{equation}
    2 \sum_{i=2}^r Re\left[\varepsilon_i \langle v_1,\tilde{Q}_{v_i}^r w_1 \rangle \right] \geq -2\frac{1}{r}\left( 1-\frac{1}{r}\right)\sum_{i=1}^r \Vert v_1 \Vert \Vert w_1 \Vert \Vert v_i \Vert^2,
\end{equation}
and 
\begin{equation}
2 \sum_{\substack{i,j=2\\ i>j}}^rRe\left[\varepsilon_i \overline{\varepsilon_j} \langle v_j,\tilde{Q}_{v_i}^r v_j \rangle \right] \geq -2\sum_{\substack{i,j=2\\ i>j}}^r\frac{1}{r}\left( 1-\frac{1}{r}\right)\Vert v_i \Vert^2 \Vert v_j \Vert^2.
\end{equation}
Thus,
\begin{subequations}
    \begin{align}
         q_{(1,1)}\left(-\frac{1}{r},C\right)&\geq \left(1-\frac{1}{r}\right)\Vert v_1 \Vert^2 \Vert w_1 \Vert^2-
\frac{r-1}{r^2}\vert \langle v_1 ,w_1 \rangle  \vert^2 +\left(1-\frac{1}{r}\right)^2\sum_{i=2}^r \Vert v_i \Vert^4\\
&-2\frac{1}{r}\left( 1-\frac{1}{r}\right)\sum_{i=1}^r \Vert v_1 \Vert \Vert w_1 \Vert \Vert v_i \Vert^2-2\sum_{\substack{i,j=2\\ i>j}}^r\frac{1}{r}\left( 1-\frac{1}{r}\right)\Vert v_i \Vert^2 \Vert v_j \Vert^2\\
&=  \left(1-\frac{1}{r}\right)^2 \left[\Vert v_1 \Vert^2 \Vert w_1 \Vert^2+\sum_{i=2}^r \Vert v_i \Vert^4 \right]+ 
        \frac{r-1}{r^2}\left( \Vert v_1 \Vert^2 \Vert w_1 \Vert^2 -\vert \langle v_1,w_1 \rangle \vert^2\right)\\
&-2\frac{1}{r}\left( 1-\frac{1}{r}\right)\sum_{i=1}^r \Vert v_1 \Vert \Vert w_1 \Vert \Vert v_i \Vert^2-  2\sum_{\substack{i,j=2\\ i>j}}^r\frac{1}{r}\left( 1-\frac{1}{r}\right)\Vert v_i \Vert^2 \Vert v_j \Vert^2 \\
&=\frac{1}{r}\left( 1-\frac{1}{r}\right)p_r(\Vert v_1 \Vert \Vert w_1 \Vert, \Vert v_2 \Vert^2, \hdots, \Vert v_r\Vert^2)+\frac{r-1}{r^2}\left( \Vert v_1 \Vert^2 \Vert w_1 \Vert^2 -\vert \langle v_1,w_1 \rangle \vert^2\right)\\
&\geq 0,
    \end{align}
\end{subequations}
     where $p_r$ is  the polynomial \eqref{PolynomialPositivity}.
\end{proof}

\appendix

\section{Example of a matrices saturating the form $q^{(n)}$}\label{appendix1}
 In this appendix, we present a family of matrices that violate the positivity of the form $q^{(n)}$, for $n$ even , $\alpha=-\frac{1}{2}-\varepsilon$, $\varepsilon>0$ and a rank $2$ matrix $C$.   Let $n$ even and consider  \begin{equation}
         C=\vert v_1\otimes \hdots \otimes v_n\rangle \langle v_1\otimes \hdots \otimes v_n \vert-\vert w_1\otimes \hdots \otimes w_n\rangle \langle w_1\otimes \hdots \otimes w_n \vert,
     \end{equation} 
     with $\Vert v_i \Vert=\Vert w_i \Vert=1$ and $v_i \perp w_i$ for every $1 \leq i \leq n$. In this case,  for $J\in P(\{1,\hdots,n\})$,
     \begin{equation}
         \Vert \tr_J C\Vert_2^2=\left\{\begin{array}{cl}
              0 & \text{if } J=\{1,\hdots,n\}  \\
              2& \text{otherwise} 
         \end{array}\right.
     \end{equation}
     so for every $\varepsilon >0$,
    \begin{subequations}
    \begin{align}
        q^{(n)}\left(-\frac{1}{2}-\varepsilon,C\right)&=2\sum_{J\in  P(\{1,\hdots,n\}) \setminus \{1,\hdots,n\}  }\left( -\frac{1}{2}-\varepsilon \right)^{\vert J\vert} \\
        &=2\sum_{k=0}^{n-1}\binom{n}{k}\left( -\frac{1}{2}-\varepsilon \right)^k\\
        &=2\sum_{k=0}^{n-1}\sum_{m=0}^k\binom{n}{k}\binom{k}{m}\frac{(-1)^k}{2^{k-m}}\varepsilon^m\\
        &=2\sum_{m=0}^{n-1}\left[ \sum_{k=m}^{n-1}\binom{n}{k}\binom{k}{m} \frac{(-1)^k}{2^{k-m}} \right] \varepsilon^m,
    \end{align}
    \end{subequations}
    where in the last step we have permuted the order of  the summations. By the following Lemma, all the even powers of $\varepsilon$ vanish and the odd are negatives. Moreover it goes to zero when $\varepsilon \to 0^+$. Consequently, $q^{(n)}\left(-\frac{1}{2}-\varepsilon,C\right)<0$, for every $\varepsilon>0$ and $\rho_{\alpha}$ is not $n$-distillable for $\alpha\in 
 \left[-1,-\frac{1}{2}\right[$,  $n$ even.

    \begin{lemma}\label{LemmaCombinatorialNumbers}
    If $n \in \mathbb{N}$ is even and $m < n$, then
    \begin{equation}
         \sum_{k=m}^{n-1}\frac{(-1)^k}{2^{k-m}} \binom{n}{k}\binom{k}{m} \left\{ \begin{array}{cl}
              =0& \text{ if } m=0 \text{ or } m \text{ is even} \\
             <0 & \text{ if } m \text{ is odd}
         \end{array}\right.
    \end{equation}
    \end{lemma}
\begin{proof}
    The proof follows from the identity 
    \begin{equation}
        \sum_{k=m}^n x^k \binom{n}{k}\binom{k}{m}=\binom{n}{m}x^m(1+x)^{n-m}.
    \end{equation}
     Evaluating in $x=-\frac{1}{2}$
    \begin{equation}
    \begin{split}
        \sum_{k=m}^{n-1}\frac{(-1)^k}{2^{k-m}} \binom{n}{k}\binom{k}{m}&=2^m\sum_{k=m}^n\left( -\frac{1}{2}\right)^k\binom{n}{k}\binom{k}{m}-\frac{(-1)^n}{2^{n-m}}\binom{n}{m} \\ &=\binom{n}{m}\left(  \frac{1}{2}\right)^{n-m}((-1)^m-(-1)^n),
    \end{split}
    \end{equation}
    and the result holds.
\end{proof}

\textbf{Acknowledgements:} I would like to thank Prof. Michael M. Wolf  for his fruitful discussions  and ideas regarding this project, also  thank to Dr. Felix Huber for his help with the bibliography and finally to all reviewers, who gave me a  great feedback. Moreover, I want to thank the funding by the Deutsche Forschungsgemeinschaft (DFG,
German Research Foundation) under Germany’s Excellence Strategy-EXC2111-390814868. 

\textbf{Conflict of interest:} The author has no conflict of interests related to this publication.

\textbf{Data availability:} Data sharing is not applicable to this article as no datasets were generated or analyzed
during the current study.

\end{document}